\documentclass[usenatbib]{emulateapj}
\usepackage{graphicx}
\usepackage[flushleft]{threeparttable}
\usepackage[usenames,dvipsnames,svgnames,table]{xcolor}
\usepackage{hyperref}
\definecolor{darkblue}{rgb}{0.0,0.0,0.3}
\hypersetup{colorlinks,breaklinks,
            linkcolor=darkblue,urlcolor=darkblue,
            anchorcolor=darkblue,citecolor=darkblue}
\usepackage{amsmath,amssymb}
\usepackage{amsmath}

\usepackage{natbib}
\begin{document}

\def\etal{et al.\ \rm}
\def\ba{\begin{eqnarray}}
\def\ea{\end{eqnarray}}
\def\etal{et al.\ \rm}
\def\Fdw{F_{\rm dw}}
\def\Tex{T_{\rm ex}}
\def\Fdis{F_{\rm dw,dis}}
\def\Fnu{F_\nu}
\def\WD{\rm WD}

\newcommand\cmtrr[1]{{\color{red}[RR: #1]}}
\newcommand\cmtla[1]{{\color{blue}[LA: #1]}}


\title{1I/2017 'Oumuamua-like Interstellar Asteroids as Possible Messengers from the Dead Stars}

\author{Roman R. Rafikov\altaffilmark{1,2}}
\altaffiltext{1}{Centre for Mathematical Sciences, Department of Applied Mathematics and Theoretical Physics, University of Cambridge, Wilberforce Road, Cambridge CB3 0WA, UK; rrr@damtp.cam.ac.uk}
\altaffiltext{2}{Institute for Advanced Study, Einstein Drive, Princeton, NJ 08540}


\begin{abstract}
Discovery of the first interstellar asteroid (ISA) --- 1I/2017 'Oumuamua --- raised natural questions regarding its origin, some related to its lack of cometary activity, suggesting refractory composition. Here we explore the possibility that 'Oumuamua-like ISAs are produced in tidal disruption events (TDEs) of refractory planetoids (asteroids, terrestrial planets, etc.) by the white dwarfs (WDs). This idea is supported by spectroscopic observations of metal-polluted WDs, indicating predominantly volatile-poor composition of accreted material. We show that such TDEs sourced by realistic planetary systems (including a population of $\gtrsim 10^3$ km planetoids and massive perturbers --- Neptune-to-Saturn mass planets) can eject up to $30\%$ of planetary mass involved in TDEs  to interstellar space. Collisional fragmentation, caused by vertical collapse of the disrupted planetoid's debris inside the WD Roche sphere, channels most of its mass into 0.1-1 km fragments, similar to 'Oumuamua. Such size spectrum of ISAs (very different from the top-heavy distributions expected in other scenarios) implies that planetary TDEs can account for a significant fraction (up to $\sim 30\%$) of the ISAs. This figure is based on existing observations of WD metal pollution, which are de-biased using realistic models of circum-WD planetary systems. Such ISAs should exhibit kinematic characteristics of old, dynamically hot Galactic populations. ISA ejection in individual planetary TDEs is highly anisotropic, resulting in large fluctuations of their space density. We also show that other ISA production channels involving stellar remnants --- direct ejection by massive planets around the WDs and supernova explosions --- have difficulty explaining 'Oumuamua-like ISAs.
\end{abstract}


\keywords{planetary systems --- minor planets, asteroids: general --- minor planets, asteroids: individual ('Oumuamua)  --- white dwarfs}


\section{Introduction.}  
\label{sect:intro}


Discovery of 1I/2017 'Oumuamua --- the first unambiguous interstellar asteroid (ISA) --- by the Pan-STARRS survey \citep{Chambers} opened a new era in our study of the planetary objects across the Galaxy. This discovery was not unexpected \citep{McGlynn}, since planet formation models naturally predict ejection of large amounts of planetary material from the outer regions of forming planetary systems to the interstellar space. However, certain characteristics of this object make its discovery rather unique. 

First, 'Oumuamua does not exhibit cometary activity \citep{Jewitt,Ye,Meech}, which is at odds with the existing  planet formation models predicting ejection of predominantly volatile-rich material from the outskirts of forming planetary systems. Given that the volatile-rich objects are also much easier to detect via their cometary activity than the asteroid-like, refractory objects \citep{Engel}, this property of 'Oumuamua represents a serious challenge to any theory of its origin. Spectroscopic observations, hinting at organically rich surface \citep{Bannister,Fitz}, make 'Oumuamua's appearance similar to volatile-rich Solar System objects that suffered long exposure to cosmic rays. However, the devolatilization of 'Oumuamua's surface cannot account for the lack of cometary activity, as Oort Cloud comets reside in identical conditions of the interstellar space for Gyrs \citep{Cuk}. For these reasons, in this work we will assume that 'Oumuamua is a {\it truly refractory object}. 

Second, kinematic properties of the 'Oumuamua are somewhat unusual. \citet{Mamajek} has shown its velocity relative to the Local Standard of Rest (LSR) to be quite low, $\approx 10$ km s$^{-1}$. If 'Oumuamua spent long time (Gyrs) in the interstellar space then it should have been dynamically heated via gravitational scattering by massive objects. One might then expect its speed relative to the LSR to be comparable to the velocity dispersion of the old Galactic populations --- M-dwarfs and white dwarfs --- which is significant ($30-40$ km s$^{-1}$). 

Third, 'Oumuamua has highly elongated shape \citep{Bolin,Meech}, with its axis ratios being possibly as high as 10:1. This may suggest that this minor object was produced in some kind of a catastrophic event \citep{Drahus}.

Fourth, the discovery of 'Oumuamua may imply that the amount of mass locked in ISAs in our Galaxy is uncomfortably high. Different authors have attempted estimating the space density $n_{\rm Ou}$ of the 'Oumuamua-like interstellar asteroids with sizes $0.1-1$ km. Based on a simple rate estimate \citet{Portegies} suggest that $n_{\rm Ou}\sim (1-7)\times 10^{14}$ pc$^{-3}$. A more careful analysis of \citet{Do} arrives at $n_{\rm Ou}\approx 2\times 10^{15}$ pc$^{-3}$. Guided by this value, as well as the estimate of the 'Oumuamua's dimensions --- mean radius $R_{\rm Ou}=40$ m ($180$ m $\times 18$ m $\times 18$ m) for albedo $p=0.2p_{0.2}$ (and bulk density of 3 g cm$^{-3}$) --- from \citet{Cuk}, we deduce the following estimate for the spatial mass density of the interstellar asteroids {\it of all sizes} in the Solar neighborhood:
\ba   
\rho_{\rm ISA}\approx 0.1~\psi_m p_{0.2}^{-1.5}n_{15}~M_\oplus~\mbox{pc}^{-3},
\label{eq:rhoOu}
\ea  
where $n_{15}\equiv n_{\rm Ou}/10^{15}$ pc$^{-3}$; we use this figure as a reference in this work (see \S \ref{sect:disc} for additional discussion). 

The factor $\psi_m$ in the expression (\ref{eq:rhoOu}) accounts for the (observationally unconstrained) size spectrum of the ISAs, and is very important \citep{Raymond}. Indeed, if the ISA size distribution is such that most of the mass is concentrated in objects much larger than 'Oumuamua (e.g. a shallow power law), then $\psi_m$ should be very large, significantly increasing $\rho_{\rm ISA}$ and bringing tension to models for the origin of this population.  

For example, power law size spectrum\footnote{Such size distribution is motivated by the characteristics of the belts of minor objects in the Solar System and current models of planetesimal formation \citep{Simon2016,Simon2017}.} $dN/dR\propto R^{-\beta}$ with $\beta<4$ and most of the mass in big objects at the upper cutoff $R_{\rm cut}$ of the size spectrum, has $\psi_m \approx (R_{\rm cut}/R_{\rm Ou})^{4-\beta}$. For a collisional equilibrium size spectrum with $\beta=3.5$ \citep{Doh} and $R_{\rm cut}\sim 10^3$ km one obtains $\psi_m\sim 10^2$, considerably boosting $\rho_{\rm ISA}$. This issue is even more severe ($\psi_m\sim 10^4$, \citealt{Raymond}) for lower values of $\beta$ found in recent studies of planetesimal formation \citep{Johansen,Simon2016,Simon2017,Schafer}, presenting yet another puzzling aspect of the 'Oumuamua's discovery.

On the other hand, if the ISA size spectrum contains most of the mass in 'Oumuamua-size objects ($0.1-1$ km), then $\psi_m\sim 1$. We will come back to this issue in \S \ref{sect:implic}-\ref{sect:refine}.

In this work we explore the possibility that 'Oumuamua-like ISAs originate in {\it tidal disruptions of the (initially bound) planetary objects by the white dwarfs} (WDs). Possibility of such tidal disruption events (TDEs) of minor planets --- {\it planetoids} --- initially orbiting WDs and scattered into the low-periastron orbits by massive planets was suggested by \citet{Debes} and \citet{Jura2003} to explain observations of atmospheric pollution with high-Z elements exhibited by a large fraction (up to $50\%$) of the WDs \citep{Farihi2016}. Remnants of these TDEs are seen as particulate and gaseous debris disks inside the Roche zones of several dozen WDs \citep{Farihi2016}. Recent discovery of disintegrating objects orbiting the WD 1145+017 \citep{Vander} strongly supports this scenario for delivering refractory material to the WDs and the reality of planetary TDEs.

\begin{table}
\caption{Meaning of some variables}
\begin{tabular}{|p{1.1cm}|p{5cm}|p{0.9cm}|}
\hline \hline
Variable & Meaning & Ref\footnotemark[1]{} \\
\hline \hline 
\\
$R_{\rm Ou}$ & Mean radius of 'Oumuamua & \S \ref{sect:intro}
\\
$R$ & Physical radius of a planetoid\footnotemark[2]{} &  \S \ref{sect:ejection}
\\ 
$R_{\rm wd}$ & White dwarf radius & \S \ref{sect:ejection}
\\ 
$R_{\rm min}(a)$ & Minimum radius of a planetoid with semi-major axis $a$, at which some post-TDE debris become unbound & (\ref{eq:a_limit}) 
\\ 
$R_{\rm min}^o$ & $R_{\rm min}(a)$ for all $a<a_o$ & (\ref{eq:Rmin}) 
\\ 
$R_{\rm cut}$ & Upper cutoff radius of a spectrum of small planetoids & (\ref{eq:dfdR}) 
\\ 
$R_{\rm lg}$ & Radius of large (planet-size) planetoids & \S \ref{sect:efficiency_comp}
\\ 
$R^{\rm max}_{\rm obs}$ & Radius of largest WD-polluting object observed so far & \S \ref{sect:obs}
\\ 
$R_f$ & Radius of a fragment produced in a TDE & (\ref{sect:size}) 
\\ 
$R^{\rm max}_f$, $R^{\rm min}_f$ & Maximum and minimum sizes of a spectrum of TDE fragments & (\ref{eq:dNfdR1}) 
\\
$D(r)$ & Size of the biggest object (fragment) surviving WD tides at distance $r$ & \S \ref{sect:size}
\\ 
\hline
$r_T$ & Tidal disruption radius & (\ref{eq:r_T})  
\\ 
$q$ & Periastron distance of the planetoid orbit, $q=a(1-e)$ & \S \ref{sect:simple_model}
\\ 
$a_{\rm min}(R)$ & Minimum semi-major axis $a$, at which a TDE involving a planetoid of size $R$ results in unbound debris & (\ref{eq:a_min})  
\\
$a_0$ & Outermost semi-major axis of volatile-poor planetoids & (\ref{eq:dfda})
\\ 
\hline
$M_{\rm wd}$ & WD mass & (\ref{eq:r_T})
\\ 
$M$ & Planetoid mass & \S \ref{sect:ejection}
\\ 
$M_{\rm sm}$ & Total mass of a belt of small ($R<R_{\rm cut}$) planetoids & \S \ref{sect:efficiency_comp}
\\ 
$M_{\rm lg}$ & Total mass of a population of large (planet-size) planetoids & \S \ref{sect:efficiency_comp}
\\ 
$M_{\rm TDE}$ & Total mass of planetoids undergoing TDEs & (\ref{eq:ejection})
\\ 
$M_{\rm ej}$ & Total mass of ejected fragments & (\ref{eq:ejection})
\\ 
$\Delta M_{\rm ISA}$ & Observed mass in ISAs per WD & (\ref{eq:MperWD})
\\ 
$\Delta M_{\rm acc}$ & Observed mass of accreted planetary material (per WD) & (\ref{eq:Macc})
\\ 
$M_{\rm obs}$ & Mass in small planetoids, which contribute to the observed WD metal pollution (theoretical analog of $\Delta M_{\rm acc}$) & (\ref{eq: Mobs})
\\ 
\hline
$E_0(a)$ & Binding energy of a planetoid & \S \ref{sect:ejection}
\\
$\Delta E$ & Characteristic energy spread of fragments produced in a TDE & (\ref{eq:DeltaE})
\\ 
\hline
$f_{\rm ej}(a,R)$ & Ejection efficiency for a planetoid of size $R$ starting at semi-major axis $a$ & (\ref{eq:ejection})
\\ 
$f_{\rm ej}^o(R)$ & Ejection efficiency for planetoids of size $R$ with a semi-major axis distribution (\ref{eq:dfda}) truncated at $a_o$ & (\ref{eq:ej_R})
\\ 
$f_{\rm ej}^{\rm sm}$ & Ejection efficiency for a belt of small planetoids & (\ref{eq:f_ej_pl})
\\ 
$f_{\rm ej}^{\rm comp}$ & Ejection efficiency for a planetary system (belt of small planetoids plus a population of large planetoids) & (\ref{eq:eff_comp})
\\ 
$\tilde f_1,\tilde f_2,\tilde f_3$ & Auxiliary functions & (\ref{eq:f_ej}), (\ref{eq:f2}), (\ref{eq:f3})
\\ 
\hline
$\rho_{\rm ISA}$ & Observed space mass density of ISAs in the Galaxy & (\ref{eq:rhoOu})
\\ 
$\rho$ & Bulk density of planetoids & (\ref{eq:r_T})
\\ 
$n_{\rm Ou}$ & Space number density of 'Oumuamua-like ISAs in the Galaxy & (\ref{eq:rhoOu})
\\ 
\hline
$\alpha$ & Power law slope of the semi-major axis distribution (\ref{eq:dfda}) & (\ref{eq:dfda})
\\
$\beta$ & Power law slope of the planetoid size distribution (\ref{eq:dfdR}) & (\ref{eq:dfdR})
\\
$p$ & ISA albedo & (\ref{eq:rhoOu})
\\ 
$\psi_m$ & Ratio of the total mass in ISAs to the mass in 'Oumuamua-sized objects & (\ref{eq:rhoOu})
\\ 
\hline   
\end{tabular}
\footnotetext[1]{Reference: equation or section number} 
\footnotetext[2]{A body that gets tidally destroyed by the WD} 
\label{table:defs}
\end{table}

TDEs of a different kind --- disruptions of initially unbound stars by the supermassive black holes --- have been invoked to explain month-long flares originating from centers of galaxies \citep{Komossa}. Theoretical studies of this phenomenon  \citep{Lacy,Rees,Phinney} suggest that up to 50\% of the original stellar material ends up on unbound trajectories following the tidal disruption. Similar processes should be occurring in TDEs of planetary objects by the WDs (modulo the fact that planetoids are initially bound), providing a way of ejecting substantial amounts of refractory mass into interstellar space and giving rise to the free-floating ISAs. 

In this study we explore different aspects of the planetary TDEs by the WDs, which are relevant for understanding the characteristics of 'Oumuamua. In \S \ref{sect:ejection} we show that TDEs of planetoids initially bound to the WDs can indeed unbind significant fraction of mass (easily $\sim 30\%$ for realistic planetary architectures) participating in the disruption event. In \S \ref{sect:size} we show that planetary TDEs can also naturally produce objects with the size of 'Oumuamua; in \S \ref{sect:refine} we examine certain dynamic aspects of the planetary TDEs and show that these catastrophic events can very efficiently convert the mass of initial planetoids into the 'Oumuamua-sized ISAs, leading to $\psi_m\sim 1$. We explore the velocity distribution of ejected ISAs and implications for the 'Oumuamua's kinematic state in \S \ref{sect:kinematics}. We then use existing observations of the WD pollution by high-Z elements to set a lower limit on the amount of refractory mass that could have been ejected into the interstellar space by the WDs (\S \ref{sect:rates}). We correct this limit for observational biases in \S \ref{sect:relate_to_obs} using realistic architectures of circum-WD planetary systems, and calculate the contribution of the planetary TDEs by the WDs to the production of the 'Oumuamua-like ISAs. We discuss our results and alternative ISA production mechanisms in \S \ref{sect:disc}, and summarize our main conclusions in \S \ref{sect:summary} (we also provide short summary of each topic at the end of each \S \ref{sect:ejection}-\ref{sect:rates}). Table \ref{table:defs} provides a key to definitions used in this work.


\section{Ejection of solids in planetary disruptions by the WDs}  
\label{sect:ejection}


We consider a planetary system orbiting a WD and consisting of a variety of components --- belts of minor objects, dwarf planets and terrestrial planets (all falling into the category of {\it planetoids}) as well as massive perturbers (Neptune-to-Jupiter mass planets) --- that survived post-main sequence (post-MS) evolution of the WD progenitor \citep{Mustill,Villaver}. We focus on the innermost $\sim 10$ AU of this system as this is roughly the distance out to which the refractory objects (asteroids, terrestrial planets, etc.) should extend in a circum-WD system. These objects, initially located interior to the iceline (at 1-3 AU) during the main sequence of the WD progenitor, move out to $\lesssim 10$ AU as a result of orbital expansion caused by mass loss during the post-MS evolution \citep{Villaver}. 

Mass loss also destabilizes planetary system \citep{Debes,Frewen,Mustill2017,Cai}, resulting in gravitational scattering of planetoids by massive perturbers. Some of these objects get scattered towards the WD on almost radial orbits, enter its Roche sphere (which extends out to $\sim 1 R_\odot$) and get tidally shredded apart. Long term secular evolution due to Lidov-Kozai effect \citep{Lidov,Kozai} driven by the binary companion provides another route for almost radial planetoid orbits \citep{PetMun,Stephan}. 

Fragments resulting in such TDEs get partly accreted by the WD (polluting its atmosphere with metals, \citealt{Jura2003}), following their circularization  \citep{Veras_shrink} and eventual depletion of the compact debris disk that forms within the WD Roche sphere \citep{Rafikov1,Rafikov2,Bochka,Metzger}. At the same time, as we show in this work, some of the fragments produced in TDE can be propelled into interstellar space, becoming ISAs.

Our picture of a planetoid TDE by the WD is based, in many respects, on the theoretical framework developed over several decades \citep{Lacy,Rees,Phinney} to describe tidal disruptions of stars by the supermassive black holes \citep{Komossa}. This analogy neglects the difference between the tidal disruption of fluid objects (stars) and asteroids, which possess internal strength, but see \S \ref{sect:size}. As we will show later (\S \ref{sect:efficiency}), the dominant contribution to the interstellar debris is produced in TDEs involving the largest planetary objects, which are gravity-dominated, just like stars.  

In the classical picture of a stellar TDE a self-gravitating object of radius $R$ and bulk density $\rho$ (and mass $M=(4\pi/3)\rho R^3$) gets tidally destroyed as it crosses the tidal disruption radius $r_T$ on approach to the central mass $M_{\rm wd}$ on its (close to) parabolic orbit. For a gravity-dominated object moving on a parabolic orbit \citet{Sridhar} give the following expression for the tidal radius:
\ba
r_T=1.05\left(\frac{M_{\rm wd}}{\rho}\right)^{1/3}\approx 1.1 R_\odot \left(\frac{M_{0.6}}{\rho_3}\right)^{1/3},
\label{eq:r_T}
\ea  
where $M_{0.6}\equiv M_{\rm wd} /(0.6~M_\odot)$, $\rho_3\equiv \rho/(3$ g cm$^{-3})$.

At the point of initial disruption (i.e. at $r_T$) a spread of specific orbital energies of order 
\ba
\Delta E=\frac{GM_{\rm wd}}{r_T^2}R\approx 1.3\times 10^{12}{\rm erg}~M_{0.6}^{1/3}\rho_3^{2/3}
\left(\frac{R}{10^3{\rm km}}\right)
\label{eq:DeltaE}
\ea
gets imparted to the resuting debris \citep{Lacy}. Beyond that point the debris travels essentially ballistically along its original orbit inside the tidal disruption sphere, largely preserving this energy spread \citep{Stone}. As the fragments dive deeper into the gravitational potential of the central mass on their way to the pericenter, tidal stresses grow and continue to destroy the largest surviving fragments. This process is partly responsible for shaping the size distribution of the resulting debris, and we study it in more detail in \S \ref{sect:size}.

Unlike the unbound stars involved in TDEs with black holes, planetary objects destroyed by the WDs are {\it initially bound}, with specific energy $E_0(a)=-GM_\star /(2a)$, where $a$ is the pre-disruption semi-major axis of the planetoid. Because of that, production of unbound fragments in TDEs is possible only when $\Delta E>|E_0|$, which requires
\ba
R>R_{\rm min}(a)=\frac{r_T^2}{2a}\approx 400~{\rm km}~a_5^{-1}\left(\frac{M_{0.6}}{\rho_3}\right)^{2/3},
\label{eq:a_limit}
\ea
where $a_5\equiv a/(5$ AU). Thus, ejection of debris into the interstellar space happens only in TDEs involving sufficiently large planetoids. Some additional ejections may be possible due to subsequent scattering of the bound debris by the planets orbiting the WD \citep{Mustill2017}, but we will not consider this channel here.

Even if $R>R_{\rm min}(a)$, only a fraction of mass of the disrupted body gets ejected, namely those fragments for which the energy boost $E$ received during the TDE exceeds $|E_0|$. To calculate this fraction we adopt a method outlined in \citet{Lodato}, which uses the fact that $R\ll r_T$ and that addition of energy to the debris in the course of TDE scales linearly with the distance $\Delta R$ from the object's center (along the direction to the central mass). As a result, energy greater than $E$ gets imparted to the mass lying further than $\Delta R=(E/\Delta E)R$ from the planetoid's center along a fixed axis. In other words, defining $dM/dE$ such that the amount of mass receiving energy boost in the interval $(E,E+dE)$ is $(dM/dE)dE$, we can write \citep{Sridhar}
\ba    
\frac{dM}{dE} &=& \frac{dM}{d\Delta R}\left(\frac{dE}{d\Delta R}\right)^{-1}=\frac{2\pi R}{\Delta E}\int\limits_{\Delta R}^{R}\rho(x)xdx
\nonumber\\
&=& \frac{\pi \rho R^3}{\Delta E}\left[1-\left(\frac{E}{\Delta E}\right)^2\right],~~~E<\Delta E,
\label{eq:dMdE}
\ea    
where in deriving the last expression we assumed {\it uniform} density for the disrupted object. Thus, the fraction of mass of the original planetoid that receives energy boost {\it above} $E$ is
\ba
f(>E) &=& M^{-1}\int\limits_E^{\Delta E}\frac{dM}{dE}dE=\tilde f_1\left(\frac{E}{\Delta E}\right),~~~{\rm where}
\nonumber\\
\tilde f_1(z)&=&\frac{1}{2}-\frac{3}{4}z +\frac{1}{4}z^3.
\label{eq:f_ej}
\ea   
This, together with equations (\ref{eq:DeltaE}) and (\ref{eq:a_limit}), implies that a mass fraction 
\ba  
f_{\rm ej}(a,R)=\frac{M_{\rm ej}}{M_{\rm TDE}}=f(>|E_0(a)|)=\tilde f_1\left(\frac{R_{\rm min}(a)}{R}\right)
\label{eq:ejection}
\ea   
(for $E/\Delta E=R_{\rm min}(a)/R<1$, or else $f_{\rm ej}(a,R)=0$) of a disrupted planetoid with radius $R$ starting at semi-major axis $a$ eventually becomes unbound, adding to the population of ISAs. Here $M_{\rm TDE}$ and $M_{\rm ej}$ represent the total mass in planetoids undergoing TDEs and total mass ejected to infinity in the process, respectively. One can see that $f_{\rm ej}\to 0$ as $R_{\rm min}(a)/R=E_0/\Delta E\to 1$; in the parabolic limit $E_0=0$ and one finds $f_{\rm ej}\to 0.5$ \citep{Lacy,Rees}.


\subsection{Efficiency of ejection.}  
\label{sect:efficiency}


Tidally disrupted planetoids should naturally have a spread of initial semi-major axes $a$, especially when one accounts for the whole population of the WDs. Thus, to properly evaluate the efficiency of ISA production in planetary TDEs one needs to account for the distribution of the planetoid semi-major axes $dN/da$. Here we use a very simple model in the form of a truncated power law:
\ba    
\frac{dN}{da}=C_a a^{1-\alpha},~~~a<a_o,
\label{eq:dfda}
\ea   
so that the "surface density" of objects undergoing TDEs $(2\pi a)^{-1}dN/da\propto a^{-\alpha}$. Here $a_o$ is the highest semi-major axis, with which refractory planetesimals can arrive into the tidal sphere of the WD. Given that we are interested in refractory objects, such as 'Oumuamua, $a_o$ can be roughly associated with the post-MS location of the former iceline in a protoplanetary disk of the WD progenitor ($a_0\sim (5-10)$ AU, depending on the progenitor mass). The pre-disruption semi-major axis $a$ of planetoids experiencing strong scattering by a giant planet at a distance $r$ can vary between $r/2$ and infinity. 

Note that the distribution (\ref{eq:dfda}) does not necessarily reflect the radial distribution of the surface density of solids in a given circum-WD planetary system, simply because the efficiency of launching planetoids into orbits leading to TDEs is a function of $a$ \citep{Frewen}. Equation (\ref{eq:dfda}) is intended to merely represent a simple model for the semi-major axis distribution of {\it planetoids ending up in TDEs}, statistically averaged over the whole population of the WDs. 

According to equation (\ref{eq:a_limit}), presence of an outer edge at $a_o$ sets a {\it lower limit} $R_{\rm min}^o$ on the size of planetoids that can have at least some of their tidal debris ejected to infinity:
\ba    
R_{\rm min}^o=R_{\rm min}(a_0) =\frac{r_T^2}{2a_o}.
\label{eq:Rmin}
\ea   
Population of objects with $R<R_{\rm min}^o$ does not produce any unbound debris. Instead, all fragments resulting from tidal disruption get eventually accreted by the WD.


\begin{figure}
\plotone{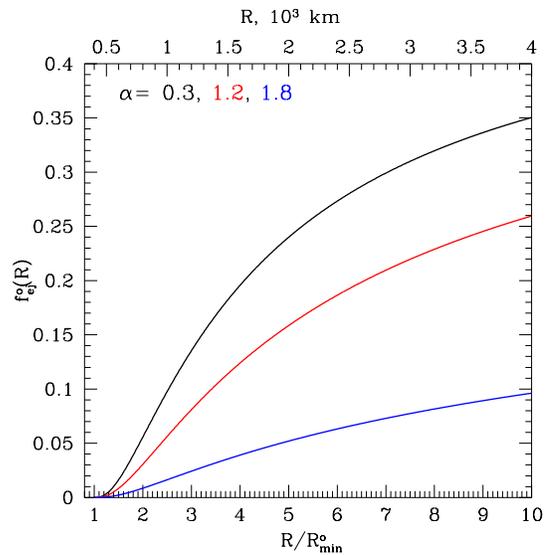}
\caption{
Fraction of mass that gets ejected to interstellar space in a tidal disruption of a planetoid of radius $R$, distributed in semi-major axis according to equation (\ref{eq:dfda}), by the WD (relative to the total planetoid mass processed in such a TDE), shown as a function of both $R/R_{\rm min}^o$ (lower axis) and $R$ (upper axis); the latter assumes $R_{\rm min}^o=400$ km as appropriate for tidal disruption of a gravity-dominated object with initial semi-major axis below $5$ AU, see equations (\ref{eq:a_limit}) \& (\ref{eq:Rmin}). Different curves correspond to different power law slopes $\alpha$ (labeled on the panel) of the spatial distribution of planetoids (\ref{eq:dfda}) with the outer cutoff at $a_o=5$ AU.
\label{fig:ej_R}}
\end{figure}


Planetoids with size $R>R_{\rm min}^o$ do produce some unbound debris, but only the objects with the pre-disruption semi-major axes $a_{\rm min}<a<a_o$, where
\ba   
a_{\rm min}(R)=\frac{r_T^2}{2R}=a_o\frac{R_{\rm min}^o}{R}.
\label{eq:a_min}  
\ea   

Based on these considerations, we can compute the efficiency $f_{\rm ej}^o(R)$ of unbound mass production in tidal disruption of objects with size $R$ and semi-major axis distribution (\ref{eq:dfda}) as 
\ba   
f_{\rm ej}^o(R) = \frac{\int_{a_{\rm min}}^{a_o}f_{\rm ej}(a,R)\frac{dN}{da} da}
{\int_0^{a_o}\frac{dN}{da} da}
=\tilde f_2\left(\frac{R}{R_{\rm min}^o}\right),
\label{eq:ej_R}
\ea   
($R>R_{\rm min}^o$) where function $\tilde f_2$ is defined by the equation (\ref{eq:f2}). Note that integration in the denominator of the equation (\ref{eq:ej_R}) runs from $a=0$ to $a=a_o$ since (in the zeroth order approximation) objects from the full range of semi-major axes can be tidally disrupted. On the other hand, for a given size $R$ production of unbound debris is possible only for $a>a_{\rm min}$, hence the integration limits in the numerator.

We plot $f_{\rm ej}^o(R)$ as a function of $R/R_{\rm min}^o$ for several values of $\alpha$ in Figure \ref{fig:ej_R}. It is clear from this figure that lower (positive) values of $\alpha$, which put more refractory mass at larger distance from the WD, result in considerably higher ejection efficiency (per unit mass in objects of a given size processed in TDEs): $f_{\rm ej}^o$ is more than 3 times higher for $\alpha=0.3$ than for $\alpha=1.8$. This is because $R_{\rm min}(a)$ is lower for more distant planetoids, driving $f_{\rm ej}^o(R)$ closer to $50\%$ for belts with more mass at high $a$, see equations (\ref{eq:f_ej})-(\ref{eq:ejection}). Also, $f_{\rm ej}^o$ grows with $R$ relatively slowly (although much faster for low values of $\alpha$), reaching $50\%$ only for very large, planetary-scale planetoids.


\begin{figure}
\plotone{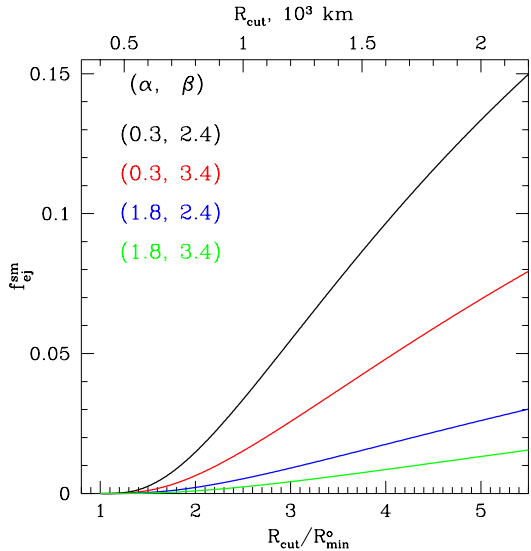}
\caption{
Fraction of mass $f_{\rm ej}^{\rm sm}$ that gets ejected to interstellar space in tidal disruptions of an ensemble of planetoids with size spectrum (\ref{eq:dfdR}) and semi-major axis distribution (\ref{eq:dfda}), relative to the total mass originating from such a planetoid population that gets processed in TDEs. This ejection efficiency is shown as a function of both the upper cutoff of the size spectrum $R_{\rm cut}$ (upper axis) as well as $R_{\rm cut}/R_{\rm min}^o$ (lower axis); analogous to Fig. \ref{fig:ej_R} the former assumes $R_{\rm min}^o=400$ km. Different curves correspond to different combinations of the power law slopes $\alpha$ and $\beta$ (labeled on the panel) characterizing the spatial and size distributions (\ref{eq:dfda}) \& (\ref{eq:dfdR}), correspondingly. Note the difference in scale of the vertical axis compared to Fig. \ref{fig:ej_R}.
\label{fig:ej_pl}}
\end{figure}



\subsection{Efficiency of ejection for a belt of planetoids with a spectrum of sizes.}  
\label{sect:efficiency_pop}

\begin{figure*}
\plotone{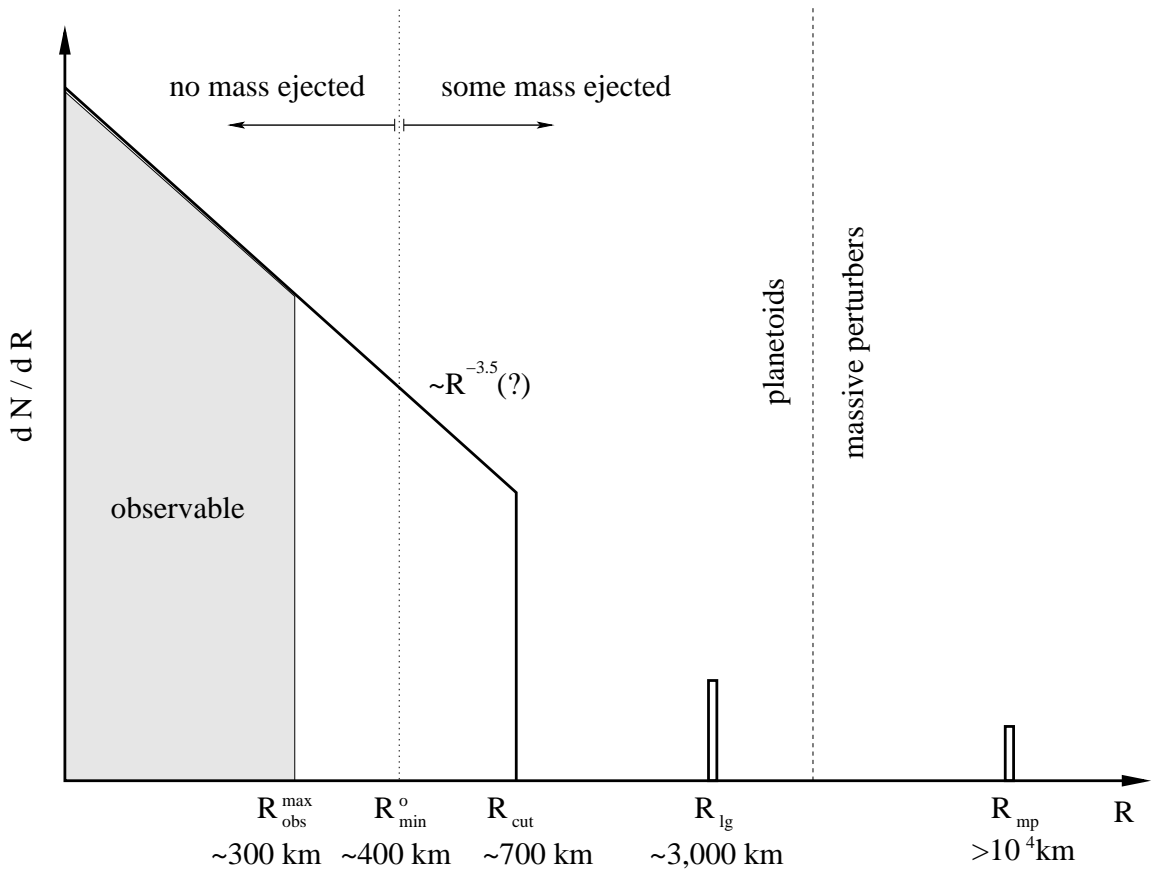}
\caption{
Schematic representation of the size distribution $dN/dR$ of a planetary system around the WD, showing its different components. Massive perturbers (to the right of dashed line) of size $R_{\rm mp}$ (ice or gas giants, stellar companions, etc.) can gravitationally drive planetoids (everything to the left of the dashed line) into the Roche sphere of the WD, leading to the TDEs. Planetoids include big objects ("planets") with radius $R_{\rm lg}$ and a belt of minor objects with a truncated size spectrum (\ref{eq:dfdR}), which extends up to $R_{\rm cut}$. Planetoids bigger than $R_{\rm min}^o$ are capable of producing some unbound fragments upon their tidal disruption by the WD; objects with $R<R_{\rm min}^o$ get fully accreted. Gray area indicates part of the size spectrum ($R<R_{\rm obs}^{\rm max}$), which contributes to the TDEs resulting in observable metal pollution of the WD atmospheres, see \S \ref{sect:relate_to_obs} for details (TDEs of bigger objects are too rare for the associated metal accretion to be caught in action). Approximate characteristic size scales are indicated on the panel. Belt size spectrum $dN/dR\propto R^{-3.5}$ (chosen simply for illustration purposes) corresponds to the Dohnanyi's collisional equilibrium \citep{Doh}.
\label{fig:spectrum}}
\end{figure*}

In practice one needs to account not only for spatial distribution of tidally destroyed objects, but also for their size distribution. We start by computing the overall efficiency of ejection produced by a belt of minor objects with a continuous mass spectrum; a more sophisticated and realistic model of a planetary system is explored next in \S \ref{sect:efficiency_comp}. We model the size distribution of planetoids undergoing TDEs as a truncated power law
\ba    
\frac{dN}{dR}=C_R R^{-\beta},~~~R<R_{\rm cut},
\label{eq:dfdR}
\ea   
where $R_{\rm cut}$ is the upper size cutoff. This model is motivated by the observed size distributions of the populations of minor objects in the Solar System --- asteroid and Kuiper belts. The belt size distribution (\ref{eq:dfdR}) is schematically illustrated in Figure \ref{fig:spectrum} with $\beta\approx 3.5$, as appropriate for the collisional equlibrium spectrum derived by \citet{Doh}.

In our calculation we again assume that planetoids of all sizes are spatially distributed according to equation (\ref{eq:dfda}). Given the finite semi-major axis extent of this distribution, production of unbound fragments in TDEs is possible only if $R_{\rm cut}>R_{\rm min}^o$ (otherwise all planetoids are accreted with $100\%$ efficiency). Whenever this condition is fulfilled, the fraction of mass ejected in TDEs marginalized over the power law distribution (\ref{eq:dfdR}) is given by
\ba
f_{\rm ej}^{\rm sm}=\frac{\int_{R_{\rm min}^o}^{R_{\rm cut}}f_{\rm ej}^o(R)\frac{dN}{dR} R^3 dR}
{\int_0^{R_{\rm cut}}\frac{dN}{dR} R^3 dR}
=\tilde f_3\left(\frac{R_{\rm cut}}{R_{\rm min}^o}\right)
\label{eq:f_ej_pl}
\ea
(note that the integration in the denominator runs between $0$ and $R_{\rm cut}$ as planetoids of all sizes can be involved in TDEs), where 
\ba   
\frac{R_{\rm cut}}{R_{\rm min}^o}\approx 2.5\left(\frac{R_{\rm cut}}{10^3~\mbox{km}}\right)\left(\frac{a_o}{5~\mbox{AU}}\right)\left(\frac{M_{0.6}}{\rho_3}\right)^{-2/3},
\label{eq:Rad_rat}
\ea  
see equations (\ref{eq:r_T}) and (\ref{eq:Rmin}). The explicit expression\footnote{We do not write down the straightforward but cumbersome analytic expression for this function.} for the function $\tilde f_3$ is given by equation (\ref{eq:f3}). 

In Figure \ref{fig:ej_pl} we plot the ejection efficiency $f_{\rm ej}^{\rm sm}$ for planetoid belt with the power law size distribution (\ref{eq:dfdR}), as a function of the upper size cutoff of the size spectrum $R_{\rm cut}$. Comparing with Figure \ref{fig:ej_R}, one can see that ejection efficiency drops significantly, when a spectrum of planetoid sizes is considered. This is because size distribution (\ref{eq:dfdR}) contains certain amount of mass in objects with sizes below $R_{\rm min}^o$, which produce no unbound fragments. Also, even for $R_{\rm cut}$ substantially higher than $R_{\rm min}^o$, belt members with $R\sim R_{\rm min}^o$ still have low ejection efficiency, driving $f_{\rm ej}^{\rm sm}$ down. 


Shallow size distributions (lower values of $\beta$) result in higher $f_{\rm ej}^{\rm sm}$, as such size spectra have more mass in largest objects with $R\sim R_{\rm cut}$, for which the ejection efficiency is highest. Also, as in Figure \ref{fig:ej_R}, higher values of $\alpha$ (i.e. less mass at large semi-major axes $a\sim a_o$) result in considerably lower ejection efficiency.


\subsection{Efficiency of ejection for a planetary system} 
\label{sect:efficiency_comp}


Finally, we evaluate the efficiency of ISA production in TDEs sourced by the more realistic planetary systems. We consider a composite model for the mean population of tidally destroyed planetoids, in which a belt of minor objects with size distribution (\ref{eq:dfdR}) is supplemented with a number of "planets" bigger than $R_{\rm cut}$, but still significantly smaller than massive perturbers that do the scattering, see Figure \ref{fig:spectrum}. This population of big objects with radius $R_{\rm lg}>R_{\rm cut}$ has total mass $M_{\rm lg}$ and we will assume that it dominates the TDE mass budget. In other words, $M_{\rm lg}\gtrsim M_{\rm sm}$, where $M_{\rm sm}$ is the total mass of minor planetoids (power law part of the size spectrum, see equation (\ref{eq:dfdR})) that end up in TDEs. At least based on the Solar System experience, in which asteroid belt co-exists with a population of terrestrial planets, one expects $M_{\rm lg}\gg M_{\rm sm}$ --- the mass of our asteroid belt is just $10^{-3}$ of the combined mass of the terrestrial planets. For simplicity we also assume the bulk density of "planets" to be the same as for small planetoids.

Given that essentially all dynamical processes emplacing planetoids onto the low angular momentum orbits (scattering by giant planets, Lidov-Kozai cycles, etc.) are insensitive to the planetoid mass, this model should provide a good description of a rather general planetary system architecture. 

We also assume that, in a statistical sense (i.e. averaged over many circum-WD planetary systems), the spatial distributions of both the high and low mass planetoids are the same and are well represented by the equation (\ref{eq:dfda}). The number of big objects per planetary system does not need to be integer and should be treated as representing the mean over many systems.

Using the results of \S \ref{sect:efficiency}-\ref{sect:efficiency_pop} we can then write the overall efficiency of mass ejection in the framework of this composite model as
\ba  
&& f_{\rm ej}^{\rm comp} = \frac{f_{\rm ej}^{\rm sm}M_{\rm sm}+f_{\rm ej}^o(R_{\rm lg})M_{\rm lg}}{M_{\rm sm}+M_{\rm lg}}
\nonumber\\
&& = \frac{M_{\rm sm}\tilde f_3\left(R_{\rm cut}/R_{\rm min}^o\right)+M_{\rm lg}\tilde f_2\left(R_{\rm lg}/R_{\rm min}^o\right)}{M_{\rm sm}+M_{\rm lg}}.
\label{eq:eff_comp}   
\ea    


\begin{figure}
\plotone{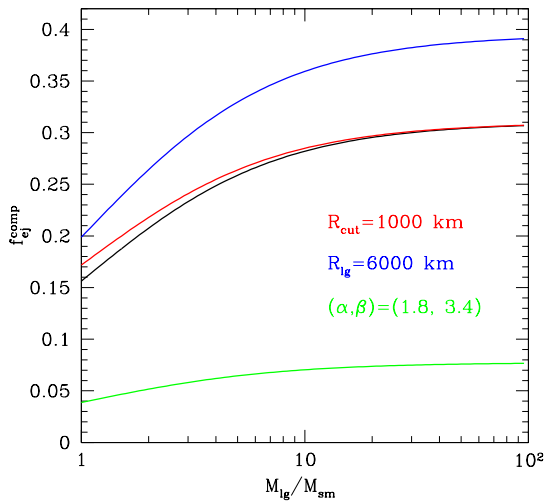}
\caption{
Fraction of mass that gets ejected to interstellar space in tidal disruptions of an ensemble of planetoids drawn from a "planetary system"-like size distribution described in \S \ref{sect:efficiency_comp} (see Fig. \ref{fig:spectrum}), shown as a function of $M_{\rm lg}/M_{\rm sm}$ --- ratio of the mass coming from the population of big objects ("planets") and from the belt of small planetoids (\ref{eq:dfdR}). Black curve is the fiducial model with $(\alpha,\beta)=(0.3,2.4)$, $R_{\rm cut}=600$ km, $R_{\rm lg}=3000$ km; $R_{\rm min}^o=400$ km is assumed throughout. Other curves show how $f_{\rm ej}^{\rm comp}$ changes when we vary just one of the parameters, as shown on the panel. 
\label{fig:ej_comp}}
\end{figure}


We plot the ejection efficiency $f_{\rm ej}^{\rm comp}$ in Figure \ref{fig:ej_comp} as a function of $M_{\rm lg}/M_{\rm sm}$ --- mass ratio of the "planetary" and belt populations. We vary different parameters of the model away from their fiducial values ($(\alpha,\beta)=(0.3,2.4)$, $R_{\rm cut}=600$ km, $R_{\rm lg}=3000$ km; $R_{\rm min}^o=400$ km) to understand how they affect the outcome. Comparing with Figure \ref{fig:ej_pl}, one can see that adding a population of massive planetoids to the planetary system considerably increases ejection efficiency. This is not surprising, since $f_{\rm ej}$ rapidly increases with the size $R$ of a disrupted object, see Figure \ref{fig:ej_R}. As a result, big planetoids completely dominate $f_{\rm ej}^{\rm comp}$ for $M_{\rm lg}/M_{\rm sm}\gtrsim 1$. This shows in Figure \ref{fig:ej_comp} as rather strong sensitivity of the ejection efficiency to the size of "planets" $R_{\rm lg}$ and only weak sensitivity to the characteristics of the belt of minor objects (e.g. $R_{\rm cut}$).

As before (see \S \ref{sect:efficiency_pop}-\ref{sect:efficiency_comp}), higher values of $\alpha$ and $\beta$ (less mass in massive, distant objects) result in considerable reduction of ejection efficiency --- even when most of the mass is in large objects ($M_{\rm lg}/M_{\rm sm}\gtrsim 10$) $f_{\rm ej}^{\rm comp}$ can barely reach $\sim 10\%$.


\subsection{Efficiency of ejection summary} 
\label{sect:efficiency_sum}


Results of this section demonstrate that tidal disruptions of planetoids by the WDs can eject substantial amounts of mass into the interstellar space. This is despite the fact that the disrupted planetoids are initially {\it gravitationally bound} to the WD, which somewhat lowers the efficiency of mass ejection in these events compared to the TDEs of initially unbound stars by the supermassive black holes \citep{Lacy,Rees,Lodato}. Nevertheless, mass-weighted ejection efficiency of $\sim 30\%$ can be easily achieved, provided that the population of disrupted objects contains most of the mass in large (several $10^3$ km) objects. Asteroid belts are far less efficient at producing unbound fragments in tidal disruptions of their members. Moreover, objects with sizes $\lesssim 400$ km starting on semi-major axes $\lesssim 5$ AU get fully accreted in TDEs, leaving no unbound fragments. 

Thus, efficient production of refractory, 'Oumuamua-like ISAs in planetary TDEs by the WDs (with efficiency of $\gtrsim 30\%$) requires that massive, planetary scale objects dominate the mass budget of planetoids involved in TDEs. This will be important when discussing observational constraints in \S \ref{sect:rates}.


\section{Size distribution of ejected fragments}  
\label{sect:size}


We now address a question of whether asteroids with the size of 'Oumuamua ($\sim 10^2$ m) should be expected to result from TDEs of planetoids by the WDs. 

Upon crossing the tidal sphere of the WD, i.e. at $r=r_T$, an incoming planetoid disintegrates into smaller objects. Results of \S \ref{sect:efficiency}-\ref{sect:efficiency_comp} suggest that fragment ejection is efficient only for large planetoids, $R\gtrsim 400$ km, which are gravity-dominated, thus we will focus on such objects. Process of their tidal disruption is akin to that of an incompressible fluid sphere, which was explored by \citet{Sridhar} among others. However, as the object loses coherence, effects of internal strength must become important at some scale. At least initially, crack propagation driven by tidal deformation will keep breaking the planetoid into ever smaller fragments, up to the point when the internal strength of these fragments becomes sufficient enough to overcome tidal stresses.   

In this picture initial fragmentation should stop when the fragment size $R_f$ becomes comparable to the maximum size $D(r_T)$ of an internally-strong object still capable of resisting tidal destruction at $r=r_T$. Given the material properties of an object, it is possible to calculate how this maximum size $D(r)$ of an object marginally surviving in the external tidal field varies with the distance to the central mass $r$. Such calculations have been attempted by a number of authors \citep{Dobrovolskis,Davidsson} and we will use their results.

However, fragmentation of the original planetoid does not stop at $r_T$. If its pericenter distance $q$ is smaller than $r_T$, then the fragments produced at $r_T$ will experience steadily increasing tidal stress. This will eventually cause them to break again until the reduction of their size restores the dominance of internal stresses over the central tides. As a result, one should expect tidal breakup of fragments into smaller ones to continue as the debris travels towards the pericenter. 

In \S \ref{sect:simple_model} we outline a very simple, "quasi-static" model for the fragment size spectrum based on these ideas with the main goal of showing that objects with the dimensions of 'Oumuamua can naturally result from tidal disruptions. We will then demonstrate in \S \ref{sect:refine} the limitations of this model and indicate qualitatively how the spectrum should be modified due to the dynamic nature of realistic tidal disruptions.


\subsection{Simple model for the size spectrum.}  
\label{sect:simple_model}


We start by adopting a simple, quasi-static picture, in which at every distance $r<r_T$ from the WD the characteristic size of of a fragment $R_f(r)$ is given by the maximum size of an object that can still sustain tidal stresses at this separation; in other words, we assume a one-to-one relation between $R_f$ and $r$ in the form $R_f(r)=D(r)$. As $r$ decreases during the debris transit through the tidal sphere, so does $D(r)$, until the pericenter at $r=q=a(1-e)$ is reached. After the pericenter passage tidal stress starts decreasing and fragmentation stops. In this picture, beyond the pericenter the TDE debris should have a size spectrum with most of the mass concentrated around $D(q)$. Variation of $q$ between $R_{\rm wd}$ and $r_T$ for different TDEs then results in spread of values of $D(q)$, shaping the size spectrum of ISAs over multiple TDEs. 

\citet{Dobrovolskis} has carried out an exploration of the breakup conditions for bodies with non-zero internal strength, the results of which were also  summarized in \citet{Davidsson} in the form of $r(D)$ relations for the minimum distance, at which objects with different material properties and size $D$ can still sustain tidal stress. In our simple model, these results allow us to {\it uniquely relate} the characteristic (mass-bearing) size of the resultant fragments to the periastron distance $q$ of the original orbit of a parent planetoid, i.e. 
\ba   
q=q(R_f),
\label{eq:rpRf}
\ea  
where we identified $R_f=D(q)$. It is important that this relation is independent of either the mass of the parent planetoid, or its original semi-major axis --- only the periastron distance matters in setting the final fragment size.

In this simple, quasi-static picture the size spectrum $dN_f/dR_f$ averaged over many TDEs can be computed as 
\ba   
\frac{dN_f}{dR_f}\approx \frac{dq}{dR_f}\int_0^\infty \left(\frac{R}{R_f}\right)^3
\frac{d^2N}{dR dq}dR,
\label{eq:dNfdR}  
\ea   
where $d^2N/dR dq$ is the distribution of tidally disrupted planetoids per unit radius $R$ and per unit periastron distance $q$, while the factor $\left(R/R_f\right)^3$ gives the number of fragments with size $R_f$ spawned by a planetoid of size $R$. 

If the reason driving planetoids into the low-periastron orbits is strong scattering by massive perturbers (e.g. a giant planet), which to zeroth order occurs roughly isotropically (i.e. with uniform distribution of the post-scattering velocity component perpendicular to the radius vector to the WD), then planetoids should have roughly uniform distribution in $l^2$ around $l=0$, where $l=\sqrt{GM_\star a(1-e^2)}$ is the specific angular momentum. This means that highly eccentric planetoids enroute to tidal disruption should have {\it roughly uniform distribution of  the periastron distance} $q$. The same conclusion was reached by \citet{Katz} in their study of secular Lidov-Kozai evolution \citep{Lidov,Kozai} leading to WD collisions in triple systems --- a completely different scenario resulting in highly eccentric orbits, which has also been invoked to explain atmospheric pollution of the WDs \citep{PetMun,Stephan}. Based on these considerations, it is legitimate to approximate  
\ba  
\frac{d^2N}{dR dq}\approx r_T^{-1}\frac{dN}{dR},
\label{eq:d2N}
\ea   
since the periastron distance varies in the range $R_{\rm wd}<q<r_T$ and $R_{\rm wd}\ll r_T$ for WDs. Using this result as well as the independence of the $q(R_f)$ relation upon the parent planetoid size, we find that, roughly, 
\ba   
\frac{dN_f}{dR_f}\approx \frac{3}{4\pi}\frac{M_{\rm TDE}}{\rho r_T}
~R_f^{-3}\frac{dq}{dR_f}
\label{eq:dNfdR1}  
\ea  
for $R_f^{\rm min}<R_f<R_f^{\rm max}$ and zero outside this range. Here $R_f^{\rm min}=R_f(R_{\rm wd})$ is the minimum fragment size, which is reached when the pericenter $q$ distance gets close to the WD surface --- at this location the tidal stress is highest and only the smallest (internally strongest) fragments can survive. Similarly, $R_f^{\rm max}=R_f(r_T)$ is the size of the largest objects that can survive tides at $r_T$ due to their internal cohesion. Also, $M_{\rm TDE}=\int_0^\infty M(dN/dR)dR$ is the total mass in planetoids that have undergone TDEs. Note that $dN_f/dR_f$ depends only on the full mass of disrupted planetoids $M_{\rm TDE}$ and not on their initial size spectrum $dN/dR$ (as long as $R\gtrsim R_f^{\rm max}$). The dependence (\ref{eq:dNfdR1}) of the size spectrum of TDE fragments (equivalent to the size spectrum of interstellar asteroids) on $R_f$ is fully contained in $R_f^{-3}dq/dR_f$. 

We now illustrate the implications of this result using a simple prescription for the material properties of a putative planetoid assumed to be composed primarily of iron. This choice does not reflect any expectations about the composition of 'Oumuamua and is selected because of associated mathematical simplicity. Qualitative picture remains the same in the case of rocky planetoids, see \S \ref{sect:rocky}.


\begin{figure*}
\plotone{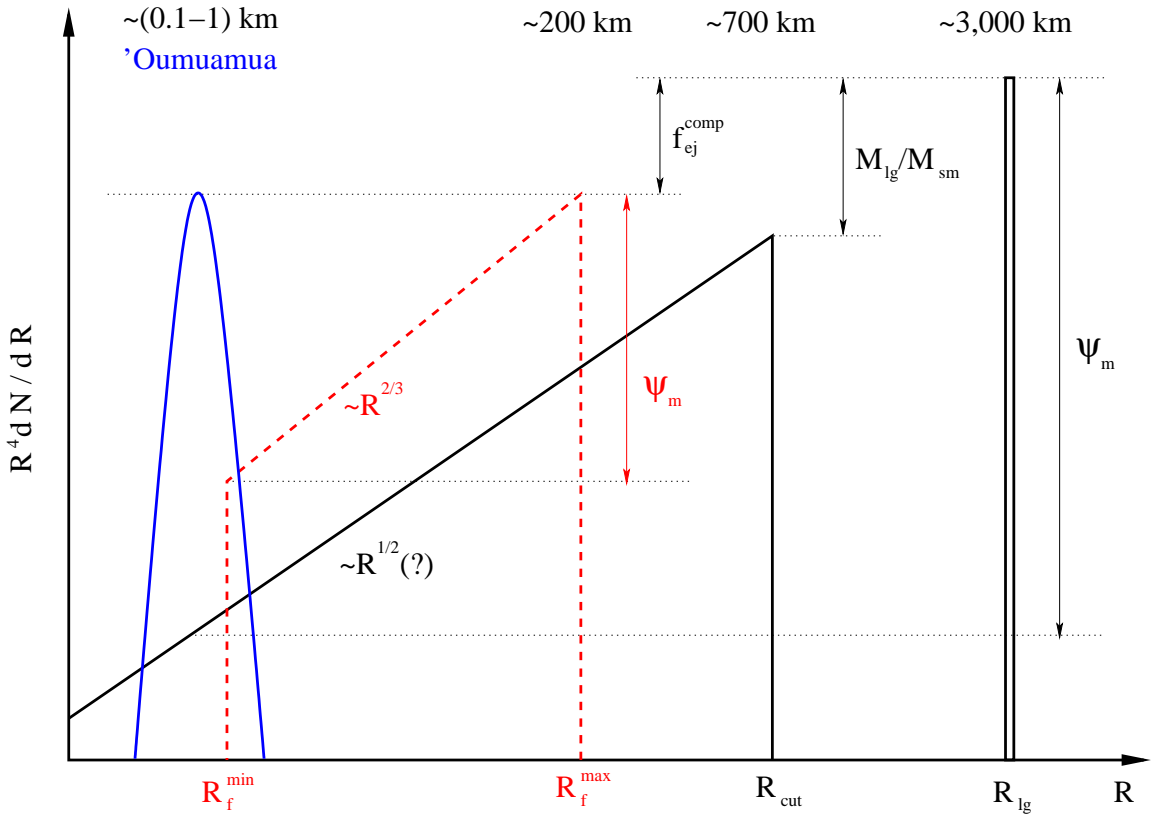}
\caption{
Schematic picture of the size distributions of the original planetoids feeding the TDEs ({\it black}, see Fig. \ref{fig:spectrum} for comparison) and resultant {\it unbound} fragments predicted by the quasi-static ({\it red}, \S \ref{sect:simple_model}) and dynamic ({\it blue}, \S \ref{sect:refine}) models. A close analogue of the cumulative mass, $R^4dN/dR$ is plotted as a function of the object size $R$ to illustrate how the mass gets transferred between different populations during the disruption event (note the reduction of mass by $f_{\rm ej}^{\rm comp}$ as planetoids get converted into unbound fragments). Original planetoid population has high value of $\psi_m$ --- ratio between the full population mass and mass in objects comparable to the 'Oumuamua --- as indicated on the panel. Quasi-static model (which is too simplistic and incomplete, thus shown with dashed curve) predicts roughly power law size spectrum of fragments extending from $\sim 100$ m to $\sim 200$ km, with slope given by equation (\ref{eq:dist_iron}) and most of the mass still in biggest fragments (again, $\psi_m\sim 10^2\gg 1$, see equation (\ref{eq:psi_simple})). A more refined dynamic TDE model, accounting for the collisional evolution caused by the vertical compression of a disrupted planetoid, puts most of the ejected mass in objects with sizes around $0.1-1$ km, close to 'Oumuamua's dimensions (resulting in $\psi_m\sim 1$).
\label{fig:mass_tran}}
\end{figure*}



\subsubsection{Iron planetoids.}  
\label{sect:iron}


For a planetoid composed primarily of iron ($\rho\approx 8$ g cm$^{-3}$) \citet{Davidsson} suggests that the tidal splitting distance, which we associate with $q$, is 
\ba   
q(R_f)\approx 0.85\left(\frac{M_{\rm wd}}{\rho}\frac{p_0(R_f)}{S}\right)^{1/3},~~{\rm iron},
\label{eq:rel1}
\ea    
where $S$ is the shear strength ($S\approx 3\times 10^3$ bar for iron) and central pressure $p_0$ is related to the fragment size $R_f$ via
\ba   
p_0(R_f)=\frac{2}{3}\pi G \rho^2 R_f^2.
\label{eq:p_0}
\ea
These relations imply that $q(R_f)\propto R_f^{2/3}$, such that 
\ba   
\frac{dN_f}{dR_f}\propto R_f^{-10/3}.
\label{eq:dist_iron}
\ea  

Setting $q=R_{\rm wd}$ in equation (\ref{eq:rel1}) we find 
\ba   
R_f^{\rm min} &\approx & 0.88\left(\frac{S}{G M_{\rm wd} \rho}\right)^{1/2}R_{\rm wd}^{3/2}
\label{eq:Rfmin_iron}\\
&\approx & 350~{\rm m}~M_{0.6}^{-0.5}\left(\frac{R_{\rm wd}}{10^{-2}R_\odot}\right)^{1.5}.
\nonumber
\ea   
This estimate shows that a $R=10^3$ km iron asteroid passing very close to the WD surface, at $q\sim R_{\rm wd}$ (which according to equation (\ref{eq:d2N}) should happen in several per cent of planetoid TDEs), will be disintegrated into at least several$\times 10^9$ fragments of size $R_f^{\rm min}$.

Next, setting $q=r_T$ we find from equation (\ref{eq:r_T})
\ba   
R_f^{\rm max} \approx  0.95\left(\frac{S}{G\rho^2}\right)^{1/2}
\approx 250~{\rm km}.
\label{eq:Rfmax_iron}
\ea   
This is the size of fragments produced in TDEs of planetoids just barely entering the Roche sphere of a WD. Figure \ref{fig:mass_tran} schematically illustrates the mass spectrum of fragments (red dashed curve) predicted by the quasi-static model for iron planetoids.


\subsubsection{Rocky planetoids.}  
\label{sect:rocky}


Disruption of rocky asteroids can also be explored using the corresponding prescription for $D(r)$ from \citet{Dobrovolskis}, which is somewhat more complicated than in the case of iron and uses tensile strength instead of shear strength. Without showing details of the calculation, we outline the main results: (1) the minimum and maximum fragment sizes are $R_f^{\rm min}\approx 100$ m and $R_f^{\rm max}\approx 200$ km; (2) the size spectrum is still reasonably well described by $dN/dR\propto R_f^{-10/3}$ with some deviations around $R_f\approx 100$ km caused by a transition in $D(r)$ behavior for rock.


\subsubsection{Implications of the simple model.}  
\label{sect:implic}


Results of \S \ref{sect:iron}-\ref{sect:rocky} demonstrate that, even in the framework of a simple model of tidal disruption outlined in \S \ref{sect:simple_model}, planetary TDEs naturally provide a possibility of producing 'Oumuamua-size objects: our estimates of $R_f^{\rm min}$ for both iron and rocky fragments are rather close to the estimated 'Oumuamua's dimensions of $\sim 10^2$ m. 
At the same time, the size spectrum of fragments predicted by the simple, quasi-static model has most of the mass in big objects, see equation (\ref{eq:dist_iron}) and Figure \ref{fig:mass_tran}. Steep fragment size distributions have been previously found in tidal disruption simulations of \citet{Debes2012}, although they had rather low resolution. Since our model also predicts $R_f^{\rm max}\gg R_f^{\rm min}$ this again raises the "large $\psi_m$" issue mentioned when discussing our estimate  (\ref{eq:rhoOu}) for $\rho_{\rm ISA}$: only a small fraction of mass in ISAs would then be contained in 'Oumuamua-like objects because $\psi_m\gg 1$ (see red $\psi_m$ symbol in Figure \ref{fig:mass_tran}). Indeed, integrating equation (\ref{eq:dNfdR1}) over the appropriate parts of the fragment size spectrum, one can estimate the ratio between the amount of mass in small objects with size $R_f^{\rm min}\sim R_{\rm Ou}$ and the full mass of ISAs contained mainly in large bodies with size $\sim R_f^{\rm max}$ as
\ba   
\psi_m\approx \frac{q(R_f^{\rm max})}{q(R_f^{\rm min})}=\frac{r_T}{R_{\rm wd}}\sim 10^2.
\label{eq:psi_simple}
\ea
Thus, this simple model for the TDE fragment mass spectrum predicts $10^2$ times more mass in ISAs than can be inferred from a single detection of 'Oumuamua.

However, as we show next, the quasi-static model is too simplistic and the estimate (\ref{eq:psi_simple}) is too high.


\subsection{Dynamical refinements to the quasi-static model.}
\label{sect:refine}


One of the most dramatic outcomes of stellar disruptions by supermassive black holes is a strong vertical (orthogonal to the midplane) compression of a star into a short-lived, dense, pancake-like configuration at some point within the Roche sphere \citep{Carter}. Vertical collapse of a star with initial radius $R_\star$ and mass $M_\star$ into such a flattened object is caused by the component of the black hole gravity normal to the orbital plane, which cannot be balanced by the pressure forces inside the Roche sphere. Prior to the maximum compression (when the density and pressure at the center become high enough to resist vertical collapse) vertical motion towards the midplane has characteristic velocity \citep{Carter,Stone}
\ba   
v_z\sim \frac{r_T}{q}\left(\frac{GM_\star}{R_\star}\right)^{1/2},
\label{eq:v_vert}
\ea   
significantly exceeding the escape velocity from the surface of the original star for deep plunges with the periastron distance $q\ll r_T$. Homologous compression at such high velocity can have dramatic consequences, including detonation of the whole star \citep{Carter}.

Our quasi-static picture of the planetoid TDE presented in \S \ref{sect:simple_model}-\ref{sect:implic} completely overlooks this highly dynamic aspect of the disruption phenomenon. An obvious difference with the stellar disruption case is the weak compressibility of a tidally destroyed planetoid. However, this is unlikely to eliminate the main features of the dynamic vertical compression picture. Indeed, to zeroth order one may consider tidally destroyed planetoid as a rubble pile composed of fragments with sizes $R_f\sim D(r)$ at each point in orbit. Uncompensated vertical gravity of the WD will initiate the collapse of individual elements of this rubble pile towards the midplane of the orbit, which is better described as a granular flow. A good analogy would be the collapse of a pyramid of billiard balls on a pool table once the rack holding its base in place is suddenly removed. 

Lack of compressibility may result in additional lateral (parallel to the orbital plane) expansion of the sheared rubble pile, but this will not prevent high-speed convergent motion of the fragments near the midplane. Moreover, conversion of the vertical kinetic energy $\sim v_z^2$ into that of the in-plane motions is unlikely to significantly affect our estimate (\ref{eq:DeltaE}) of the energy spread $\Delta E$. Indeed, using equations (\ref{eq:r_T}), (\ref{eq:DeltaE}), and (\ref{eq:v_vert}) one can show that $v_z^2/\Delta E\sim r_T R/q^2$, where in (\ref{eq:v_vert}) we replaced $R_\star$, $M_\star$ with $R$, $M$. As a result, vertical energy can exceed $\Delta E$ only for $q\lesssim (r_T R)^{1/2}$, which is $\lesssim 0.1r_T$ even for largest planetoids; i.e. $v_z^2\gtrsim \Delta E$ in at most several per cent of TDEs, given the flat distribution (\ref{eq:d2N}) of $q$. However, even in these rare events most of the vertical energy will still be dissipated inelastically (rather than contribute to in-plane motions) as we describe now.

Head-on collisions of fragments comprising the two hemi-spheres of the original planetoid, which are vertically collapsing towards the orbital midplane at speeds indicated by the equation (\ref{eq:v_vert}), will result in catastrophic destruction of debris objects and fragmentation cascade down to small sizes. Indeed, even if equation (\ref{eq:v_vert}) overestimates the vertical velocity by a factor of several, $v_z$ is still so high that it would easily exceed the escape speed  (which scales $\propto R_f$) from the surface of any fragment by a considerable margin, leading to catastrophic disruptions. 

This short-duration burst of fragmentation will likely cease, or at least slow down, when the fragment size becomes small enough for the material strength to start opposing the collisional splitting (since the minimum velocity necessary for catastrophic disruption increases with decreasing size in the strength-dominated regime, \citealt{Housen}). In the absence of a good model of either the granular flow-like vertical collapse or the ensuing fragmentation cascade, we hypothesize that the final size distribution of fragments resulting from a planetoid TDE should be {\it peaked around the size at which the collision velocity leading to catastrophic disruption is minimized}. Using the results of \citet{Stewart}, this final size of the surviving fragments can be estimated as \citep{RS15}
\ba   
R_f^{\rm dyn}\sim (0.1-1)~{\rm km}, 
\label{eq:min_size}
\ea
depending on the material properties of colliding objects. 

Needless to say, this estimate is rather uncertain, even though it is supported by physical arguments. A more accurate calculation of $R_f^{\rm dyn}$ should be possible via direct simulations of tidal disruptions of asteroid-like objects \citep{Debes2012,Movsho,Veras_TDE}. Such simulations could (1) show whether the overall dynamic picture as presented here is correct, (2) determine the magnitude of the convergent vertical motions deep inside the Roche sphere of the WD for comparison with equation (\ref{eq:v_vert}), and (3) provide input for calculating collisional evolution of individual "rubble particles" comprising tidally disrupted planetoid with the ultimate goal of determining the characteristic fragment size $R_f^{\rm dyn}$ resulting from a TDE.

In the dynamic picture of the planetary TDE presented here, most of the planetoid mass gets converted into objects with rather narrow spread in sizes around $\sim R_f^{\rm dyn}$ (schematically represented by the blue curve in Figure \ref{fig:mass_tran}). Given that $\sim R_f^{\rm dyn}$ is very close to the dimensions of 'Oumuamua, this implies that the resultant size spectrum of unbound fragments would have 
\ba   
\psi_m\sim 1. 
\label{eq:dyn_psi}
\ea   
As a result, the estimate (\ref{eq:rhoOu}) should then properly reflect the spatial mass density of ISAs in the Galaxy. 

Highly dynamic nature of the disruption event, followed by substantial collisional evolution of its products, may also provide important clues for understanding the highly elongated shape and the rotation state of the 'Oumuamua \citep{Drahus}.


\subsection{Size distribution summary} 
\label{sect:size_sum}


Results of this section demonstrate that planetary TDEs should be efficient at producing fragments with sizes comparable to the 'Oumuamua's dimensions. Even the simple quasi-static model presented in \S \ref{sect:simple_model} successfully produces objects with sizes in the  $0.1-1$ km range, although the overall amount of mass ending up in such fragments is predicted to be rather small, around $1\%$ (formally resulting in high $\psi_m\sim 10^2$).

A more refined dynamic model (\S \ref{sect:refine}), accounting for the vertical collapse of a rubble pile, into which the planetoid turns inside the WD Roche sphere, predicts substantial collisional grinding of the TDE products (which may also be relevant for explaining the unusual shape and rotation of the 'Oumuamua). We hypothesize that collisional evolution caused by convergent vertical motions will convert most of the original planetoid mass into objects with sizes around $0.1-1$ km (see Figure \ref{fig:mass_tran}). Below this size scale internal strength of fragments becomes important, making further collisional fragmentation of the disruption products difficult. 

A very important outcome of this dynamic planetary TDE model is the expectation of $\psi_m\sim 1$ for the size distribution of fragments that it predicts. The implications of this prediction will be further discussed in \S \ref{sect:rates}, when comparing the existing observational constraints on the metal accretion by the WDs with the inferred space density of the ISAs.


\section{Kinematic properties of ejected fragments}  
\label{sect:kinematics}


Unbound ejecta leaving the potential well of the WD have a range of velocities at infinity $v_\infty$. For an object with a pre-disruption radius $R$ and semi-major axis $a$ receiving energy boost $E>|E_0(a)|$ energy conservation dictates that 
\ba   
\frac{v_\infty^2}{2}=E+E_0(a)=\Delta E \left[\frac{E}{\Delta E}-\frac{R_{\rm min}(a)}{R}\right], 
\label{eq:v_infty}
\ea
see equations (\ref{eq:DeltaE})-(\ref{eq:a_limit}). The mass-weighted distribution of $v_\infty$, i.e. the {\it mass fraction} $dm/dv_\infty$ of ejecta with velocity in the interval $(v_\infty,v_\infty+dv_\infty)$, is easily found using equations (\ref{eq:dMdE})  and (\ref{eq:v_infty}) to be 
\ba
&& \frac{dm}{dv_\infty} =  \frac{v_\infty}{f_{\rm ej}(a,R)M}\frac{dM}{dE}
\nonumber\\
&& =  \frac{3}{4f_{\rm ej}(a,R)}\frac{v_\infty}{v_0^2}\left[1-\left(\frac{v_\infty^2}{2v_0^2}+\frac{R_{\rm min}(a)}{R}\right)^2\right],
\label{eq:dmdv}
\ea   
with $f_{\rm ej}(a,R)$ given by equation (\ref{eq:ejection}) and 
\ba  
v_0=\left(\Delta E\right)^{1/2}\approx 12~{\rm km~s}^{-1}M_{0.6}^{1/6}
\rho_3^{1/3}
\left(\frac{R}{10^3{\rm km}}\right)^{1/2}.
\label{eq:v0}
\ea   
Distribution (\ref{eq:dmdv}) has a finite extent in the velocity space,
\ba   
v_\infty<v_{\rm max}=2^{1/2}v_0\sqrt{1-\frac{R_{\rm min}(a)}{R}},
\label{eq:vmax}
\ea   
for $R>R_{\rm min}(a)$, and is normalized to unity. Note that equation (\ref{eq:dmdv}) does not assume any {\it size distribution} of the unbound fragments; it simply characterizes the total amount of mass locked in all debris objects moving with certain velocity $v_\infty$. 


\begin{figure}
\plotone{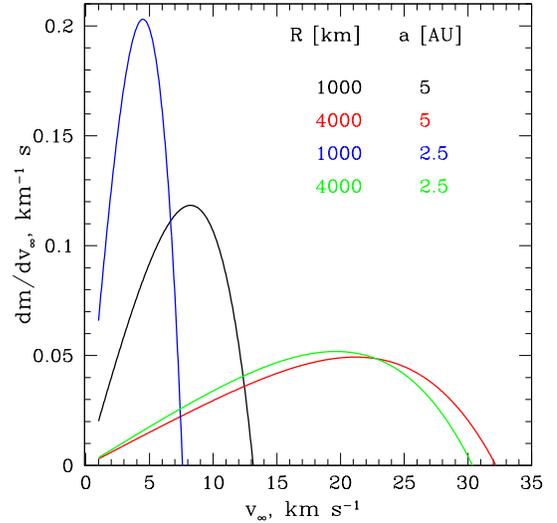}
\caption{
Velocity distribution function (mass-weighted and normalized to unity) of unbound fragments resulting from tidal disruption of a planetoid with initial radius $R$ and semi-major axis $a$. Different curves correspond to different pairs of $R$ and $a$, as shown on the panel. 
\label{fig:vel_dist}}
\end{figure}


In Figure \ref{fig:vel_dist} we plot $dm/dv_\infty$ for several values of $R$ and $a$ of a parent object. One can see that this velocity distribution is somewhat weighted towards higher values of $v_\infty$. For bigger disrupted objects $dm/dv_\infty$ extends to higher velocities, as expected from equations (\ref{eq:v0})-(\ref{eq:vmax}). Decreasing the semi-major axis of the parent object results in lower ejection speeds, in accordance with equations (\ref{eq:a_limit}), (\ref{eq:vmax}). However, the sensitivity to $a$ is far less pronounced when tidal disruption involves a big object, as $R_{\rm min}(a)/R$ is smaller in this case. Roughly speaking, unbound fragments produced in planetoid TDEs should be ejected with velocities of order $10-30$ km s$^{-1}$.

A notable feature of the material ejection in TDEs is that it is highly anisotropic, with the unbound fragments leaving the WD in the form of a narrow jet. This jet has a substantial internal velocity shear along its direction ${\bf n}_{\rm ej}$ ($|{\bf n}_{\rm ej}|=1$), with speeds ranging from $0$ to $v_{\rm max}$. Velocity of each ejected fragment gets summed up with the speed ${\bf v}_{\rm wd}$ of the WD motion in the Galaxy, so that the initial speeds of unbound fragments ${\bf v}_f$ can be described as a one-parametric family
\ba   
{\bf v}_f={\bf v}_{\rm wd}+\lambda v_{\rm max}{\bf n}_{\rm ej}
\label{eq:1par}
\ea    
with $0<\lambda<1$. The small natural width\footnote{This width can be increased by the post-disruption interaction of the debris with the circum-WD planets  \citep{Mustill2017}.} of the velocity distribution of the original jet of fragments is unimportant compared to the spread of initial velocities given by the equation (\ref{eq:1par}). As a result, every planetary TDE releases a {\it narrow filament} of objects at a point in a Galaxy, with objects moving in different directions having quite different (but uniquely related to the ${\bf v}_f$ direction) speeds. This motion is very different from a homologous, quasi-spherical expansion. 

Equation (\ref{eq:1par}) implies different, but highly clustered in phase space, initial conditions for the subsequent motion of fragments in the Galactic potential. This likely means that full spatial mixing of the debris from a single TDE throughout the Galaxy should take a long time, since the ejected objects will tend to remain close to certain 1-dimensional manifolds (continuously stretching filaments) determined by their initial ejection conditions. As a result, spatial distribution of the ejected fragments inside the Galaxy can be rather {\it spatially clustered}, especially for relatively young ISAs, further complicating the determination of their mean Galactic density (see \S \ref{sect:disc} for further discussion). 

Averaging over the ensemble of planetary TDEs produced by different WDs one can, of course, assume that the directions of the jets of ejecta are random. The determination of the full velocity distribution of the debris fragments with respect to the Local Standard of Rest (LSR) requires additional convolution of equation (\ref{eq:dmdv}) with the velocity distribution of the WDs, which can be modeled as a triaxial velocity ellipsoid. Dimensions of the WD velocity ellipsoid vary depending on the age of the WD, reflecting the history of their dynamical excitation by the molecular clouds and transient spiral arms  \citep{BT}. For hot ($T>12,000$ K), young WDs \citet{Elsan} found $(\sigma_R,\sigma_\phi,\sigma_z)=(29,18,23)$ km s$^{-1}$ for the components of the velocity ellipsoid, while for cold ($T<12,000$ K), old WDs they cite $(\sigma_R,\sigma_\phi,\sigma_z)=(42,31,36)$ km s$^{-1}$ (although these samples are very different in size). Given the typical scale of the WD ejection velocity (see equation (\ref{eq:v0})), it is likely that the overall velocity distribution of the ISAs in the Galaxy is only slightly dynamically hotter than that of the WDs from which they originated (which is already quite hot). 

After the ISAs are released from the WD, their random velocities with respect to the LSR get further pumped up by the same mechanisms that dynamically heat old stellar populations --- gravitational scattering by massive perturbers and transient density waves. Thus, if the debris resides in the Galactic disk for Gyrs, its velocity dispersion should be close to that of the old stellar populations (WDs, M dwarfs) and rather high, around $30-40$ km s$^{-1}$.


\subsection{Implications for the observed kinematic state of 'Oumuamua} \label{sect:kinematics_impl}


\citet{Mamajek} found that prior to its passage through the inner Solar System 'Oumuamua had rather low velocity relative to the LSR, around $10$ km s$^{-1}$ in the azimuthal direction, with negligible vertical and radial Galactocentric velocities. This motion appears somewhat unusual in light of the expectations outlined above, which suggest that ISA velocities of up to $30-40$ km s$^{-1}$ in the LSR frame may be typical. At the same time, for a population with a given velocity ellipsoid the probability of drawing an object with a particular velocity relative to the LSR peaks at zero velocity. 

Nevertheless, the observed kinematic state of 'Oumuamua would be more likely if it originated from a population with low velocity dispersion. This observation may hold important clues for the origin of this ISA, since it would imply that this object (1) has originated from a dynamically cold Galactic source and (2) is relatively young. Regarding the latter, if 'Oumuamua had spent long ($\gtrsim$ Gyr) time unbound in the Galaxy, its epicyclic motion would be excited by random gravitational perturbation to levels (on average) exceeding its measured space motion. However, if its youth is the consequence of its short lifetime (e.g. against fading or some other decay mechanism) then the production rate of the ISAs should be increased correspondingly \citep{Gaidos} to maintain their observed abundance $\rho_{\rm ISA}$ given by equation (\ref{eq:rhoOu}). This is likely to be challenging. A similar possibility is that we are catching 'Oumuamua at the very beginning of its long life in the interstellar space, which is, again, a low probability event. 

In light of these arguments, we prefer to think that 'Oumuamua's slow Galactic motion simply reflects the peak probability of drawing an object from a Schwarzschild velocity distribution in the LSR frame \citep{BT}, and that this ISA does in fact belong to the old Galactic population. Any model for its origin, in which 'Oumuamua spends $\gtrsim$ Gyr in the interstellar space, should arrive at this conclusion. Our idea of the planetary TDE is no exception in this regard.


\subsection{Kinematic properties summary} \label{sect:kinematics_sum}


Results of this section demonstrate that the unbound fragments resulting from planetary TDEs get ejected with characteristic speeds of 10-30 km s$^{-1}$, depending primarily on the mass of a disrupted planetoid, with more massive objects producing faster ejecta. The mass-weighted velocity distribution of the ejecta (\ref{eq:dmdv}) is found to be slightly weighted towards higher velocities. It should be convolved with the velocity distribution of the parent WDs (which are dynamically hot) to give full representation of the ejecta kinematics in the Galactic frame. As a result, we deduce that Gyr-old ISAs should have velocity dispersion $\sim (30-40)$ km s$^{-1}$ with respect to the LSR, similar to the old stellar populations.  

In light of these findings, we interpret the observed low velocity of 'Oumuamua with respect to the LSR as simply reflecting the highest probability of picking an object from a Schwarzschild velocity distribution. Subsequent detections of the ISAs should find a significant fraction of them to be fast, dynamically-hot objects. This conclusion should hold regardless of the origin mechanism of the ISAs, as long as they spend long time in the interstellar space.  

ISA ejection in TDEs is highly anisotropic and clustered in phase space, leading to slow phase mixing of the ejecta across the Galaxy. As a result, spatial distribution of the ISAs can exhibit significant fluctuations in density.


\section{Constraints on the abundance of ISAs based on observations of metal-accreting WDs}
\label{sect:rates}


Since WDs represent the key element of our model for the production of 'Oumuamua-like objects, it is natural to convert the estimate (\ref{eq:rhoOu}) into the amount of mass in ISAs {\it per WD}.

\citet{Fukugita} estimate the space mass density of the WDs in the local neighborhood as $(5.5\pm3.0)\times 10^{-3}M_\odot$ pc$^{-3}$. Assuming a mean WD mass of $0.6M_\odot$, this translates into the WD space number density of $\approx 10^{-2}$ pc$^{-3}$. Equation (\ref{eq:rhoOu}) then implies that our Galaxy on average contains 
\ba
\Delta M_{\rm ISA}\approx 10~\psi_m p_{0.2}^{-1.5}n_{15}~M_\oplus~~\mbox{per WD}
\label{eq:MperWD}
\ea
in ISAs.  
  
We now use observational constraints on the accretion of refractory material by the WDs and the results of previous sections to see whether the ejection of fragments of the planetary TDEs by the WDs can account for the amount of mass in the refractory interstellar asteroids (per WD) implied by the detection of 'Oumuamua.


\subsection{Observational constraints on the amount of accreted solids}  
\label{sect:obs}


A very fortunate feature of planetary material processing by the WDs is that we can actually characterize it observationally. Atmospheres of tens of per cent of all WDs are known to be polluted with metals \citep{Farihi2016}, and the most plausible mechanism for this pollution is accretion of the planetary material in the TDEs \citep{Debes,Jura2003}, an idea strongly supported by the recent discovery of the disintegrating objects near the WD 1145+017 \citep{Vander}. The amount of mass accreted by an average WD during its lifetime can be calculated if one (1) is able to measure the accretion rate $\dot M_Z$ of refractory material by the WDs and (2) has some idea of how this rate varies as the WD ages. 

\citet{Wyatt2014} used an unbiased sample of spectroscopically observed WDs to infer the statistics of accretion rates in different evolutionary stages. Since their sample is relatively large (hundreds of WDs) and unbiased, the observed instantaneous values of $\dot M_Z$ are expected to reflect reasonably well the time-averaged picture of metal accretion by the WDs (but see below for caveats). We use their results for the cumulative distribution of $\dot M_Z$ for different sub-samples to arrive at an estimate of the mean $\dot M_Z\approx 5\times 10^8$ g s$^{-1}$ for all WDs with ages of $100-500$ Myr. It should be mentioned that the determination of this mean value is a non-trivial procedure riddled with various observational biases \citep{Farihi2016}.

On the other hand, theoretical calculations of the long-term dynamics of planetary delivery into the tidal radius of the WD \citep{Mustill2017}, supported by observations \citep{Hollands2018}, suggest that $\dot M_Z$ decays exponentially with a characteristic timescale of $\approx 1$ Gyr. Normalizing this relation to the results of \citet{Wyatt2014} in the 100-500 Myr age bin and integrating $\dot M_Z$ over the lifetime of the WD (several Gy), we arrive at the following {\it observationally motivated} estimate of the full (integrated over lifetime) accreted planetary mass of
\ba   
\Delta M_{\rm acc}\sim 0.003~M_\oplus~~\mbox{per WD}.
\label{eq:Macc}
\ea   
All this mass must have been accreted in the TDEs of a kind that we envisage in this paper. A similar, although less sophisticated exercise has been carried in \citet{Hansen}, who arrived at an estimate of $\Delta M_{\rm acc}$, which is 3 times higher than (\ref{eq:Macc}), primarily because they assumed $\dot M_Z$ to remain at a constant level of $3\times 10^8$ g s$^{-1}$ over 5 Gyr. 

The estimate (\ref{eq:Macc}) is orders of magnitude lower than the mass in interstellar asteroids per WD given by equation (\ref{eq:MperWD}). At face value, this discrepancy represents a serious blow to the idea that planetary tidal disruptions by the WDs can provide a significant source of interstellar asteroids. Indeed, results of \S \ref{sect:efficiency} imply that mass ejected in TDEs should be further lowered compared to $\Delta M_{\rm acc}$ by a factor of $\bar f_{\rm ej}/(1-\bar f_{\rm ej})\sim 0.4$ (for the mean ejection efficiency\footnote{Scattering of initially bound debris by circum-WD planets \citep{Mustill2017} can additionally increase $\bar f_{\rm ej}$.} $f_{\rm ej}=0.3$), meaning that only $\bar f_{\rm ej}(1-\bar f_{\rm ej})^{-1}\Delta M_{\rm acc}/\Delta M_{\rm Ou}\approx 0.01\%$ of interstellar asteroids can be explained by tidal disruptions of planetoids by the WDs.

However, there are strong reasons to believe that the estimate (\ref{eq:Macc}) likely represents a {\it lower limit} on $\Delta M_{\rm acc}$ and the actual accreted mass is much higher. Indeed, statistics of observed $\dot M_Z$ in \citet{Wyatt2014} is heavily dominated by the WDs with the highest values of $\dot M_Z$, around $10^{10}$ g s$^{-1}$. Also, we know a number of metal-rich WDs, which accrete at even higher rates, in excess of $10^{11}$ g s$^{-1}$ \citep{Bergfors}. Adding just several such objects to the sample of \citet{Wyatt2014} would increase the estimate (\ref{eq:Macc}) by an order of magnitude. Thus, the determination of $\Delta M_{\rm acc}$ is heavily affected by the small number statistics at the highest $\dot M_Z$.

What this means is that current observations of $\dot M_Z$ are likely revealing to us only the very tip of an iceberg\footnote{We are in disagreement on this point with \citet{Hansen}, who assumed that measurements of the WD metal pollution are essentially complete.}, and the highest $\dot M_Z$ events resulting from the accretion of largest ("planet"-scale) planetoids that dominate the budget of accreted mass are so rare that our current WD sample is simply too small to catch them in action. 

The maximum size of a planetoid that is sampled by current observations of $\dot M_Z$ can be estimated by measuring the total mass in refractory elements contained in the outer convective envelopes of actively accreting WDs. According to \citet{Farihi2010}, the largest amounts of atmospheric Ca measured in a handful of objects correspond to accreted mass equivalent to the mass of a $R=R_{\rm obs}^{\rm max}\approx 300$ km asteroid (similar to Vesta). This can be considered as the {\it maximum size of objects, to which our measurements of $\dot M_Z$ are sensitive} because of the rarity of accretion events involving even more massive objects (which at the same time deliver more mass to the WD, boosting $\Delta M_{\rm acc}$). Note, that $R_{\rm obs}^{\rm max}$ is below our estimate (\ref{eq:a_limit}) for $R_{\rm min}^o$ (see Figure \ref{fig:spectrum}), meaning that the currently observed $\dot M_Z$ in metal-rich WDs derive from planetary TDEs, which did not eject any material into the interstellar space. 



\subsection{Correspondence between the observationally inferred accreted mass and the mass in unbound fragments}  
\label{sect:relate_to_obs}


We now describe a (model-dependent) procedure, by which we account for the biases mentioned in \S \ref{sect:obs} to allow a more meaningful relation to be established between the observed $\Delta M_{\rm acc}$ and the ejected mass $M_{\rm ej}$.

As in \S \ref{sect:efficiency_comp}, we assume that the planetoid population destined for TDEs can be represented, in a statistically averaged sense, as a combination of a belt of minor planets (asteroids) with the size spectrum given by the equation (\ref{eq:dfdR}) and total mass $M_{\rm sm}$ plus a population of large objects with size $R_{\rm lg}>R_{\rm cut}$ and total mass $M_{\rm lg}$. 

As we saw in \S \ref{sect:obs}, so far observations of the WD metal pollution have been telling us only about the properties of relatively small objects.  For that reason we will assume that planetoids with size $\lesssim R_{\rm obs}^{\rm max}$ belong to the belt of minor planets, i.e. $R_{\rm obs}^{\rm max}<R_{\rm cut}$. In other words, the manifestations of TDEs involving only a part of the full mass spectrum (illustrated with the grey region in Figure \ref{fig:spectrum}) are observable at present. 

The total "observable" mass contained in this part of the size spectrum is
\ba   
M_{\rm obs}=M_{\rm sm}\left(\frac{R_{\rm obs}^{\rm max}}{R_{\rm cut}}\right)^{4-\beta}<M_{\rm sm}. 
\label{eq: Mobs}  
\ea    
We will also assume that the ejection efficiency is zero for $R<R_{\rm obs}^{\rm max}$, i.e. that the "observable" planetoids are small enough to get fully accreted. This is consistent with our estimate $R_{\rm obs}^{\rm max}\approx 300$ km $<R_{\rm min}^o$ and equations (\ref{eq:a_limit}) \& (\ref{eq:Rmin}).

Using the results of \S \ref{sect:efficiency_comp} we can then estimate the ratio between the true ejected mass $M_{\rm ej}$ and the observable mass $M_{\rm obs}$ for such a system as
\ba   
\frac{M_{\rm ej}}{M_{\rm obs}}
& = & \frac{f_{\rm ej}^{\rm comp} \left(M_{\rm lg}+M_{\rm sm}\right)}{M_{\rm obs}}=\left(\frac{R_{\rm cut}}{R_{\rm obs}^{\rm max}} \right)^{4-\beta}
\nonumber\\
& \times & \left[\frac{M_{\rm lg}}{M_{\rm sm}}\tilde f_2\left(\frac{R_{\rm lg}}{R_{\rm min}^o}\right)+\tilde f_3\left(\frac{R_{\rm cut}}{R_{\rm min}^o}\right)\right],
\label{eq:ej_vs_acc}   
\ea   
where we used equation (\ref{eq: Mobs}) and $f_{\rm ej}^{\rm comp}$ given by equation (\ref{eq:eff_comp}). 


\begin{figure}
\plotone{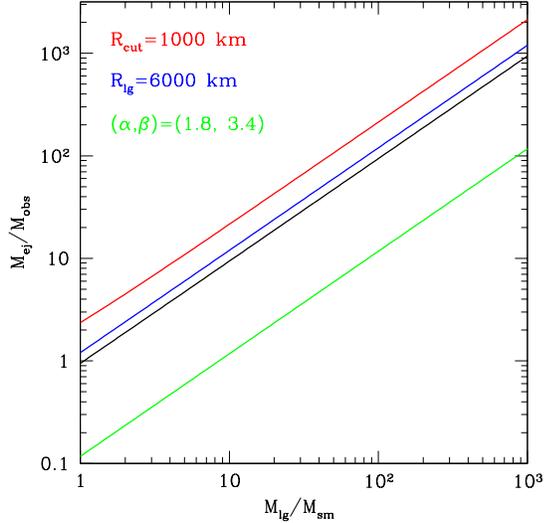}
\caption{
Ratio between the mass in ISAs ejected through planetary TDEs ($M_{\rm ej}$) and the fraction of mass accreted by the WD, which is potentially accessible to current spectroscopic measurements of metal pollution ($M_{\rm obs}$). Shown as a function of $M_{\rm lg}/M_{\rm sm}$ --- ratio of the mass processed in TDEs originating from the population of big objects ($M_{\rm lg}$) and from the power-law asteroid belt population ($M_{\rm sm}$) (\ref{eq:dfdR}). Black curve is the fiducial model with $(\alpha,\beta)=(0.3,2.4)$, $R_{\rm cut}=600$ km, $R_{\rm lg}=3000$ km; $R_{\rm min}^o=400$ km is assumed throughout. Other curves show how $f_{\rm ej}^{\rm comp}$ changes when we vary just one of the parameters, as shown on the panel. One can see that $M_{\rm ej}/M_{\rm obs}$ can easily reach $10^2$ for $M_{\rm lg}/M_{\rm sm}\gtrsim 10^2$.
\label{fig:Mej_Mobs}}
\end{figure}


In Figure \ref{fig:Mej_Mobs} we plot $M_{\rm ej}/M_{\rm obs}$ for fixed $R_{\rm obs}^{\rm max}= 300$ km, $R_{\rm min}^o=400$ km (corresponding to $a_0=5$ AU, see equation (\ref{eq:Rmin})) and different values of the upper cutoff $R_{\rm cut}$, radius of the planetary-scale planetoids $R_{\rm lg}$ and the mass ratio between the populations of "planets" and minor objects $M_{\rm lg}/M_{\rm sm}$. One can see that the mass ejected in TDEs involving planetary systems with massive objects ("planets") can easily exceed the mass in metals probed by the WD spectroscopic observations by $\gtrsim 10^2$, provided that that total mass in these big bodies processed through TDEs is $\gtrsim 10^2$ times larger than the mass in small, asteroid-like objects. For reference, masses of the asteroid belt and terrestrial planets in the inner Solar System are such that $M_{\rm lg}/M_{\rm sm}\sim 10^3$.

This result is relatively insensitive to either the size of the big objects, or the upper cutoff of the spectrum of the asteroid-like objects. It is quite sensitive to both spatial and size distributions of the disrupted objects (indices $\alpha$ and $\beta$). However, even for the rather unfavorable case ($\alpha=1.8$, $\beta=3.4$) shown in Figure \ref{fig:Mej_Mobs}, ejected mass still exceeds the observable mass by $\sim 10$ when the big, thousand-km objects dominate mass of the planetary system ($M_{\rm lg}/M_{\rm sm}\gtrsim 10^2$). 

These considerations allow us to estimate the relative contribution (by mass) of the planetary TDEs to the observed ISA population as 
\ba   
\frac{M_{\rm ej}}{\Delta M_{\rm ISA}} &=& \frac{M_{\rm ej}/M_{\rm obs}}{\Delta M_{\rm ISA}/\Delta M_{\rm acc}}
\nonumber\\
&\approx & 0.3~\psi_m^{-1} p_{0.2}^{1.5}n_{15}~\frac{M_{\rm ej}/M_{\rm obs}}{10^3},
\label{eq:rel_abund}
\ea   
where we identified $M_{\rm obs}$ with $\Delta M_{\rm acc}$ and used estimates (\ref{eq:MperWD}) and (\ref{eq:Macc}). Results of \S \ref{sect:size} suggest that $\psi_m\sim 1$ for the population of ISAs produced in planetary TDEs, which is a critical ingredient. Then, equation (\ref{eq:rel_abund}) implies that such events can potentially account for a significant share --- up to $\sim 30\%$ --- of 'Oumuamua-like ISAs, provided that (1) 'Oumuamua's albedo is not much lower than 0.2 and (2) metal accretion by the WDs is dominated on average by big, planetary scale (several $10^3$ km) planetoids, and that $M_{\rm ej}/M_{\rm obs}\gtrsim 10^3$. 

The latter implies a significant abundance of planetary-scale planetoids, which raises an issue of the observability of the consequences of their TDEs. To assess this question we compare the number of such planetoids $N(R_{\rm lg})$ with the number of planetoids $N(\sim R_{\rm obs}^{\rm max})$ of radius comparable to the maximum "observed" size $R_{\rm obs}^{\rm max}$ in our composite model of the planetary system, see \S \ref{sect:efficiency_comp}. Assuming that mass spectrum of small planetoids (\ref{eq:dfdR}) is dominated by the largest objects, i.e. $\beta<4$, we can relate $N(\sim R_{\rm obs}^{\rm max})$ to the number of planetoids near the upper size cutoff $N(\sim R_{\rm cut})$ as $N(\sim R_{\rm obs}^{\rm max})/N(\sim R_{\rm cut})\approx \left(R_{\rm obs}^{\rm max}/R_{\rm cut}\right)^{1-\beta}$. Since also $N(R_{\rm lg})/N(\sim R_{\rm cut})\approx (M_{\rm lg}/M_{\rm sm})\left(R_{\rm cut}/R_{\rm lg}\right)^3$, one can relate
\ba
&& \frac{N(R_{\rm lg})}{N(\sim R_{\rm obs}^{\rm max})}\approx \frac{M_{\rm lg}}{M_{\rm sm}}\left(\frac{R_{\rm cut}}{R_{\rm lg}}\right)^3\left(\frac{R_{\rm obs}^{\rm max}}{R_{\rm cut}}\right)^{\beta-1}
\nonumber\\
&& \approx  0.4~\frac{M_{\rm lg}/M_{\rm sm}}{10^3}
\left(\frac{5}{R_{\rm lg}/R_{\rm cut}}\right)^{3}\left(\frac{10^3\mbox{km}}{R_{\rm cut}}\right)^{2.5},
\label{eq:num_rat}
\ea  
where we set $\beta=3.5$ and $R_{\rm obs}^{\rm max}=300$ km.

This estimate demonstrates that the number of large planetoids can be comparable to the number of largest objects that are currently observed to be engulfed by the WDs (for $M_{\rm lg}/M_{\rm sm}=10^3$, which according to Figure \ref{fig:Mej_Mobs} and equation (\ref{eq:rel_abund}) is needed for planetary TDEs to contribute substantially to the population of ISAs). Given that we have seen $\sim 1$ of the latter kind of objects, it is plausible that the expansion of the spectroscopically-studied sample of the WDs by a factor of several may result in a detection of a system that underwent a TDE involving an Earth-sized object relatively recently (although the proper interpretation of the nature of such an object may be an issue, given the high metal abundance in its atmosphere).


\subsection{ISA abundance summary}  
\label{sect:abundance_sum}


Spectroscopic observations of metal-polluted WDs suggest that, on average, they accrete $\Delta M_{\rm acc}\sim 0.003 M_\oplus$ of refractory material over their lifetime. However, this estimate is based on observations of systems that have accreted only relatively small, asteroid-like objects in recent past. It completely misses the much larger amount of mass infrequently delivered by big, planetary-scale objects: such events are too rare to be observed at the moment given our limited sample of objects. 

We correct for this observational bias by considering a simple model for the architecture of a typical planetary system feeding the WD accretion, and demonstrate that the estimate (\ref{eq:Macc}) should be boosted by a factor $M_{\rm ej}/M_{\rm obs}\sim 10-10^3$, provided that the mass accreted in planetary objects ($\gtrsim 3000$ km) exceeds that in asteroid-like bodies by a factor $\gtrsim 10-10^3$. 

Including this bias, our calculations show that planetary TDEs can potentially account for tens of per cent of 'Oumuamua-like ISAs, provided that (1) $\psi_m\sim 1$ (as suggested by the results of \S \ref{sect:size}) and (2) the accretion of refractory elements by the WDs is dominated by tidal disruptions of large, planetary scale objects (with $M_{\rm lg}/M_{\rm sm}\sim 10^3$, like in the Solar System).


\section{Discussion}  
\label{sect:disc}


This work provides analysis of different aspects of the planetary tidal disruptions by the WDs in light of the discovery of the first ISA. Here we provide additional discussion of some relevant issues. 

We believe that the (unexpected) refractory appearance of the 'Oumuamua strongly motivates planetary TDEs by the WDs as the source of the ISAs; other proposed potential sources favor production of the  volatile-rich, cometary ISAs. First and foremost, spectroscopic observations strongly suggest that, with rare exceptions \citep{Xu2017}, planetoids involved in the TDEs by the WDs and responsible for their observed metal pollution are {\it predominantly rocky and volatile-poor} \citep{Klein,JuraXu,Zuckerman}. Since the same events also contribute to the production of unbound ISAs, the planetary TDE connection appears very natural\footnote{Possible link between 'Oumuamua-like ISAs and planetary systems around the WDs was independently suggested by \citet{Hansen}, although in a different context, see \S \ref{sect:direct_ej}}. 

Second, orbital expansion of the planetoid orbits driven by the stellar mass loss during the post-MS evolution should push the refractory material, initially located interior to the iceline at 1-3 AU, further from the star. As we have shown in \S \ref{sect:ejection}, this considerably enhances production of unbound refractory fragments in planetary TDEs. Orbital expansion also naturally destabilizes circum-WD planetary systems, giving rise to planetary scattering and TDEs \citep{Debes,Debes2012}.

Third, post-MS evolution provides an opportunity for the volatile depletion of planetoids, although detailed calculations \citep{JuraXu2010,Malamud1,Malamud2} show this effect to be less dramatic than one could expect. Additional devolatilization may occur during the energetic process of tidal disruption of a planetoid, see \S \ref{sect:refine}.

Other aspects of the planetary TDEs explored in this work seem to generally support the idea that these events can indeed provide the source of refractory ISAs. Indeed, as shown in \S \ref{sect:size}, such TDEs can naturally put most of the planetoid mass in unbound objects with the size comparable to that of 'Oumuamua. This outcome logically leads to $\psi_m\sim 1$, alleviating possible tension for the spatial mass density $\rho_{\rm ISA}$ and the ISA production efficiency.

Needless to say, the ISA abundance estimate itself  (\ref{eq:rhoOu}) is very uncertain as it is based on a statistics of a single object \citep{Do}. There are many observational biases that one needs to account for in deriving this figure \citep{Engel}, which may not be fully understood. Also, $\rho_{\rm ISA}$ estimate uses certain assumptions about the physical properties of the object --- albedo, density --- which are poorly constrained. While albedo may be inferred from the future IR observations of 'Oumuamua \citep{Trilling}, the bulk density is more difficult to constrain, especially in light of the relatively slow rotation of the object. Another potential issue is the spatial inhomogeneity of ISAs in the Galaxy, naturally resulting from their anisotropic ejection in the planetary TDEs (see \S \ref{sect:kinematics}). This could mean that the estimate (\ref{eq:rhoOu}) is not representative of the true mean Galactic value of $\rho_{\rm ISA}$. Future detections of the ISAs will reveal to us how many of them could indeed be produced in planetary TDEs.

Our scenario does require processing of a significant amount of mass in large refractory objects through planetary TDEs (several $M_\oplus$ per WD, see equation (\ref{eq:MperWD})) to account for a fraction (tens of per cent) of the ISAs. This implies that substantial reservoirs ($\gtrsim 10 M_\oplus$) of refractory mass must be orbiting the WDs for this to work. At present we do not have direct observational evidence that would strongly support that possibility. However, the majority of the WDs are descendants of stars more massive than the Sun \citep{Kalirai}, and it is believed that planet formation might be considerably more efficient around such stars \citep{Johnson2007,Johnson2010}; this could quite plausibly be the consequence of their more massive refractory mass reservoirs.

It is important to emphasize that objects ending up in TDEs cannot come from the innermost parts of the circum-WD planetary systems as all objects inside $\sim 1$ AU would be engulfed during the post-MS evolution of the progenitor star \citep{Mustill,Villaver}. Instead, these objects should come from larger distances, in agreement with the considerations in \S \ref{sect:ejection}.

\citet{Frewen} explored the fates of planetoids with initial semi-major axes in the range $\sim (2-8)$ AU perturbed by planets of different masses and eccentricities. They found that Neptune-mass perturbers are far more efficient at driving planetoids into the WD Roche sphere than giant planets: the former can easily drive more mass into orbits leading to TDEs than gets directly ejected from the system (without closely approaching the WD). They also showed that eccentricity of the massive perturbers plays a key role, with highly eccentric ($e\gtrsim 0.8$), Neptune-mass planets giving rise to $>10$ times more orbits leading to TDEs than to direct ejections from the same population of planetoids. Juputer-mass objects are more efficient (by a factor $\sim 10$ for $e=0.2$) at directly scattering planetoids out of the system rather than driving them into the Roche zone of the WD; however, at higher planetary eccentricities ($e=0.8$) the difference shrinks to less than a factor of $3$. 

\citet{Mustill2017} studied the evolution of planetoids initially at $5-10$ AU perturbed by systems of three planets of different masses. Similar to \citet{Frewen} they found that lower mass planets are very efficient at driving planetoids into the low-periastron orbits: in their simulations involving Neptune and Saturn-mass planets about $30\%$ of planetoids entered the Roche sphere of the WD (and super-Earths may work even better). Systems of Jupiter-like planets were more efficient at direct ejections of planetoids, with $\lesssim 10\%$ of them ending up in TDEs.

Based on these dynamical results we conclude that circum-WD planetary systems harboring planetoid belts extending out to $\sim 10$ AU with embedded Neptune-to-Saturn mass perturbers are fully capable of losing tens of per cent of their mass to TDEs, especially if these perturbers are eccentric. This should be sufficient to satisfy the constraint (\ref{eq:MperWD}) provided that $\psi_m\sim 1$, which is expected in our picture of the ISA production.


\subsection{Other ideas for the origin of 'Oumuamua}  
\label{sect:other_ideas}


We now discuss other ideas proposed for the origin of 'Oumuamua-like ISAs. Many of these suggestions \citep{Batygin,Portegies,Trilling,Raymond} appeal to {\it direct ejections} of large amounts of mass from the forming planetary systems as a result of gravitational scattering by massive perturbers (giant planets). 

There are two obvious issues with this suggestion. First, the non-volatile appearance of 'Oumuamua restricts its origin to within a few AU from the star, inside the iceline. It is well known (e.g.  \citealt{Petrovich,Morrison}) that directly scattering planetoids into interstellar space from this range of separations is extremely difficult, even for very massive planets. A lot of planetary material can be ejected from the outer regions \citep{Raymond2012}, but it will be volatile-rich, unlike 'Oumuamua.

Second, direct ejections by giant planets unbind planetoids with equal efficiency regardless of their size. As a result, in this scenario the {\it size spectrum of ISAs should directly reflect the size spectrum of the planetoids} in the parent planetary system. This immediately brings back the large $\psi_m$ issue discussed in \S \ref{sect:intro}: both the observed architecture of our Solar System and the current simulations of planetesimal formation \citep{Simon2016,Simon2017,Schafer} finding top-heavy planetoid size distributions with $\beta\approx 2.8$ suggest that $\psi_m$ should be very large, $\gtrsim 10^2$, and maybe as high as $10^4$ (\citealt{Raymond}, see black $\psi_m$ label in our Figure \ref{fig:mass_tran}). The resultant estimate (\ref{eq:rhoOu}) of the ISA abundance would then run into conflict with the availability of refractory mass in forming planetary systems and require unrealistic efficiency of ejecting planetoids. This issue cannot be overcome by adopting a bottom-heavy  size distribution of planetoids with most mass in $\sim 1$ km objects, as collisional evolution would wipe out this population on very short timescale while the WD progenitor is still on the main sequence \citep{Wyatt2007}, resulting in excessive infrared dust luminosities.

Regarding direct ejections and the large $\psi_m$ issue, it is interesting to highlight the existing constraints on the abundance of volatile-rich interstellar comets. \citet{Jura2011} used the upper limits on H abundances in the atmospheres of two He-dominated WDs to argue that the spatial density of O locked in interstellar comets (which could deliver water to these WDs) is below $\sim 0.5M_\oplus$ pc$^{-3}$, regardless of comet's size. On the other hand, \citet{Engel} set a constraint on the abundance of $\sim 1$ km interstellar comets, which is 100 times more stringent than the limit on ISAs (as the former are much easier to detect). 

These limits, combined with the constraint (\ref{eq:rhoOu}), make rather plausible the possibility that (1) the space number density of refractory objects is much higher than that of interstellar comets in the size range $\sim (0.1-1)$ km, while (2) the total (integrated over the full size spectrum) space mass density of ISAs is lower than that of interstellar comets. This situation would be a natural consequence of the {\it two different production channels} for these populations: interstellar comets can be sourced by direct ejections from the outer regions of the planetary systems and have $\psi_m\gg 1$, while ISAs are produced in planetary TDEs by the WDs with $\psi_m\sim 1$, as envisaged in this work.

\citet{Cuk} suggested a scenario, in which refractory 'Oumuamua-like objects are first produced in tidal disruptions of planetoids by dense, compact stars (red dwarfs), and then propelled to interstellar space by their binary companions. While this scenario bears some resemblance to what we consider in this work, we find it rather unlikely. First, there is no natural reason for producing large fluxes of low-periastron objects (sourcing TDEs) around red dwarfs, which are stable over many billions of years. On the contrary, in the WD case the mass loss during the post-MS evolution of the progenitor naturally results in the dynamical destabilization of the planetary system \citep{Debes}. Second, the necessity of having a companion star in a particular range of separations must significantly reduce the efficiency of this ISA production channel. And the direct ejections of fragments in the process of TDEs, i.e. without the assistance of the binary companions (as relied upon in our work) would be inefficient in the red dwarf case, since their icelines lie very close to the star.

\citet{Jackson} came up with a scenario, in which a central binary ejects {\it circumbinary} minor objects that migrate towards it through the protoplanetary disk. Given that circumbinary planets (and planetary systems) are expected to be rather common \citep{Armstrong,Silsbee}, this scenario appears rather attractive\footnote{Although the effect of the gaseous component of a circumbinary disk may significantly impact the ejection efficiency of minor objects by binaries.}. Unfortunately, like many other ideas, this one is affected by the aforementioned large $\psi_m$ issue. \citet{Jackson} also find that in their model volatile and non-volatile objects are ejected in roughly equal numbers, which seems to be in tension with the observational constraints of \citet{Engel}.


\subsection{Other ideas: direct ejections during post-MS evolution}
\label{sect:direct_ej}


Speaking of dead stars, we could not overlook other ISA production scenarios involving stellar remnants, which do not rely on unbound fragment production in planetary TDEs. 

\citet{Raymond} mention the possibility of exoplanetary Oort Cloud ejections during the post-MS evolution of the intermediate mass stars \citep{Veras2014,Stone}. However, all the released objects will be volatile-rich and would not contribute to the population of 'Oumuamua-like objects.

\citet{Hansen} suggested another obvious mechanism for injecting refractory minor objects into the interstellar space: {\it direct ejection of asteroid-like bodies} orbiting WDs by massive perturbers {\it without entering the Roche sphere} of the WD. Indeed, the same massive planets that scatter planetoids into the low-periastron orbits resulting in TDEs should also be capable of scattering planetoids into hyperbolic orbits. Moreover, the efficiency of this process can easily be much higher than scattering into orbits resulting in TDEs, especially for massive ($>1 M_J$) perturbing planets orbiting on low-eccentricity orbits \citep{Frewen}. Even for low (Neptune) mass planets \citet{Frewen} and \citet{Mustill2017} find that direct ejections unbind at least as much mass as gets processed in TDEs.

It is clear that this channel of ISA production should indeed be operating as suggested by \citet{Hansen}. However, as with other direct ejection scenarios discussed in \S \ref{sect:other_ideas}, the large $\psi_m$ problem will have its ruinous effect. Even though direct ejections of objects interior of the (radially expanded) iceline can unbind large amount of refractory mass in objects of all sizes, only a tiny fraction of this mass will be in 'Oumuamua-like objects. Thus, processing even small mass of planetoids through TDEs, which results in a population of fragments with $\psi_m\sim 1$, should still be much more efficient at producing 'Oumuamua-like ISAs.


\subsection{Interstellar asteroids resulting from core collapse supernova explosions}  
\label{sect:SNe}


Supernova (SN) explosions represent another (evolved) candidate for ejecting large amounts of refractory material into the interstellar space. Sudden loss of more than half of the initial mass during the SN explosion should unbind all orbiting planetary objects, including refractory asteroids. 

Based on nucleosynthesis arguments \citet{Arnett} estimate the number of SNe that took place in the Galaxy over its age to be $\sim 10^9$. \citet{Fukugita} provide a similar figure for the number of neutron stars and black holes in our Galaxy, based on the IMF and stellar evolution considerations. This is about a factor of 10 less than the number of WDs believed to reside in the Galaxy. Combined with equation (\ref{eq:MperWD}), these estimates suggest that to explain the abundance of the ISAs implied by the 'Oumuamua's detection, each SN explosion should be releasing  $\approx 100\psi_m M_\oplus$ in refractory objects (for 'Oumuamua's albedo of 0.2). 

Taken at face value, this amount of ejected mass may not look problematic, as one might expect massive stars ending their lives as SNe to harbor more massive planetary systems than the less massive WD progenitors (although \citealt{Reffert} found that planetary occurence drops for stars more masive than $3M_\odot$). Given that Solar System contains at least several $M_\oplus$ in refractory elements (in the form of asteroids, terrestrial planets, and, partly, cores of giants), it may not be surprising that a 20 $M_\odot$ star could be capable of releasing $\sim 10^2M_\oplus$ of refractory elements when it goes off as a SN.

However, once again, this scenario suffers from the large $\psi_m$ issue, as the ejection of planetary objects during the SN is size-independent, making the ejected mass budget prohibitive. Moreover, it is not even clear if objects in the 'Oumuamua size range ($0.1-1$ km) survive the explosion. Indeed, ejection of $10 M_\odot$ envelope by the SN would blast a 1 km asteroid orbiting at 5 AU with roughly its own mass of ejected gas moving at thousands of km s$^{-1}$. This can easily result in serious structural damage to such an unlucky asteroid (and could endow its remains with a velocity with respect to the LSR, which is too high).


\section{Summary}  
\label{sect:summary}


In this work we explored the possibility of producing interstellar asteroids in the course of tidal disruption events involving a WD and an initially bound non-volatile planetoid. This ISA formation channel is motivated by the recent discovery of the interstellar asteroid 'Oumuamua, which appears to be a refractory object. We explore different aspects of the planetary TDE --- its efficiency of generating unbound objects, their size spectrum and kinematic properties, as well as the compatibility of the observed ISA abundances with independent observational constraints (spectroscopic observations of metal pollution of the WDs). Below we briefly summarize our conclusions (more focused and extended summaries can be found at the end of each of the sections \S \ref{sect:ejection}-\ref{sect:rates}).

\begin{itemize}
    
\item Planetary TDEs by the WDs naturally explain the refractory appearance of 'Oumuamua, since spectroscopic observations of metal-polluted WDs hint at primarily volatile-poor composition of the accreted planetary material.

\item Estimate of the spatial density of the ISAs based on 'Oumuamua's detection very sensitively depends on their size spectrum. Top-heavy spectra extending over several decades in size beyond the 'Oumuamua size (as predicted by planetesimal formation models and the Solar System experience) result in unrealistically large amounts of mass locked in ISAs. This issue can be resolved if significant fraction of the interstellar refractory mass is concentrated in $\sim (0.1-1)$ km objects.

\item Efficiency with which planetary TDEs unbind mass initially locked up in tidally disrupted planetoids is a sensitive function of the planetoid size (\S \ref{sect:efficiency}). Objects with radii $\lesssim 400$ km get fully accreted by the WDs, without ejecting unbound fragments. Planetoids bigger than $10^3$ km can have $\gtrsim 10\%$ of their original mass launched onto hyperbolic orbits. Averaged over the realistic planetary system architecture (including planet-size planetoids), ejection efficiency can reach $\gtrsim 30\%$ (\S \ref{sect:efficiency_comp}).
    
\item Fragmentation during a TDE naturally channels mass of the original planetoid into significantly smaller fragments (\S \ref{sect:size}). Large convergent vertical motions arising during the planetary passage through the Roche sphere of the WD result in additional collisional grinding of fragments (\S \ref{sect:refine}). We hypothesize that this process leaves most of the original planetoid mass in $\sim (0.1-1)$ km objects (size scale where internal strength starts to dominate cohesive properties of the fragments). This alleviates the refractory ISA mass budget problem alluded to before. Dynamic nature of the TDE can also be relevant for explaining the highly elongated shape of the 'Oumuamua. 

\item Refractory ISAs ejected in planetary TDEs by the WDs should have Galactic kinematic properties similar to the old stellar populations --- M-dwarfs and WDs (\S \ref{sect:kinematics}). 'Oumuamua's slow Galactic motion can be interpreted as simply corresponding to the peak of the velocity distribution function.  
    
\item Spectroscopic observations of metal-polluted WDs provide an independent (although observationally biased) way of constraining the amount of planetary material processed in TDEs (\S \ref{sect:obs}). After accounting for the realistic architecture of the planetary systems feeding WD accretion (\S \ref{sect:relate_to_obs}), we find it quite plausible that TDEs involving large, planetary scale objects could produce {\it a significant fraction (up to tens of per cent)} of the 'Oumuamua-like objects in the Galaxy, if several $M_\oplus$ of refractory mass can be processed in TDEs by each WD.

\item Belts of refractory planetoids extending out to $\sim 10$ AU (as a result of orbital expansion during the post-MS evolution of the WD progenitor) with embedded perturbing planets in the Neptune-to-Saturn mass range (multiple or on eccentric orbits) provide natural environment for producing planetary TDEs of a kind explored in this work (\S \ref{sect:disc}).

\item Alternative explanations for the origin of  'Oumuamua must address the mass budget issue (\S \ref{sect:other_ideas}). We show that neither direct ejection of refractory planetoids by massive perturbers orbiting the WDs (tapping into the same mass reservoir that feeds the TDEs explored in this work) nor the release of planetary material during the SN explosions pass this test (\S \ref{sect:direct_ej}-\ref{sect:SNe}).
    
\end{itemize}

In linking 'Oumuamua to the planetary TDEs by the WDs our logic was often guided by the assumption of its refractory nature. If it will be shown later that this ISA is, in fact, volatile-rich, then many of the constraints used in our work would be alleviated. This, however, will not change the validity of the overall picture of the planetary TDEs by the WDs (which we know must take place for other reasons) presented here.

Upcoming IR observations of the 'Oumuamua \citep{Trilling} should help constrain its albedo, size, and spatial mass density of ISAs, see equation (\ref{eq:rhoOu}). Theoretical and numerical work on modeling tidal disruptions of rubble piles \citep{Movsho,Veras_TDE} can significantly improve our understanding of the size distribution of fragments resulting from planetary TDEs. Future detections of other ISAs by the time-domain surveys and subsequent characterization of their physical properties, size distribution and kinematic state will eventually reveal to us their true origin, the ideas for which are currently based on a single object. 

\acknowledgements

I acknowledge illuminating exchanges with Eugene Churazov, Evgeny Derishev, Matija Cuk, Michele Bannister, Jay Farihi, Brad Hansen and Brian Metzger. I am grateful to the referee (Alexander Mustill) for a thorough reading of the paper and many useful suggestion. Financial support for this study has been provided by NSF via grant AST-1409524 and NASA via grant 15-XRP15-2-0139.



\bibliographystyle{apj}
\bibliography{references}

\begin{thebibliography}{}
\expandafter\ifx\csname natexlab\endcsname\relax\def\natexlab#1{#1}\fi

\bibitem[{{Armstrong} {et~al.}(2014){Armstrong}, {Osborn}, {Brown}, {Faedi},
  {G{\'o}mez Maqueo Chew}, {Martin}, {Pollacco}, \& {Udry}}]{Armstrong}
{Armstrong}, D.~J., {Osborn}, H.~P., {Brown}, D.~J.~A., {et~al.} 2014, \mnras,
  444, 1873

\bibitem[{{Arnett} {et~al.}(1989){Arnett}, {Schramm}, \& {Truran}}]{Arnett}
{Arnett}, W.~D., {Schramm}, D.~N., \& {Truran}, J.~W. 1989, \apjl, 339, L25

\bibitem[{{Bannister} {et~al.}(2017){Bannister}, {Schwamb}, {Fraser},
  {Marsset}, {Fitzsimmons}, {Benecchi}, {Lacerda}, {Pike}, {Kavelaars},
  {Smith}, {Stewart}, {Wang}, \& {Lehner}}]{Bannister}
{Bannister}, M.~T., {Schwamb}, M.~E., {Fraser}, W.~C., {et~al.} 2017, \apjl,
  851, L38

\bibitem[{{Bergfors} {et~al.}(2014){Bergfors}, {Farihi}, {Dufour}, \&
  {Rocchetto}}]{Bergfors}
{Bergfors}, C., {Farihi}, J., {Dufour}, P., \& {Rocchetto}, M. 2014, \mnras,
  444, 2147

\bibitem[{{Binney} \& {Tremaine}(2008)}]{BT}
{Binney}, J., \& {Tremaine}, S. 2008, {Galactic Dynamics: Second Edition}
  (Princeton University Press)

\bibitem[{{Bochkarev} \& {Rafikov}(2011)}]{Bochka}
{Bochkarev}, K.~V., \& {Rafikov}, R.~R. 2011, \apj, 741, 36

\bibitem[{{Bolin} {et~al.}(2018){Bolin}, {Weaver}, {Fernandez}, {Lisse},
  {Huppenkothen}, {Jones}, {Juri{\'c}}, {Moeyens}, {Schambeau}, {Slater},
  {Ivezi{\'c}}, \& {Connolly}}]{Bolin}
{Bolin}, B.~T., {Weaver}, H.~A., {Fernandez}, Y.~R., {et~al.} 2018, \apjl, 852,
  L2

\bibitem[{{Caiazzo} \& {Heyl}(2017)}]{Cai}
{Caiazzo}, I., \& {Heyl}, J.~S. 2017, \mnras, 469, 2750

\bibitem[{{Carter} \& {Luminet}(1983)}]{Carter}
{Carter}, B., \& {Luminet}, J.-P. 1983, \aap, 121, 97

\bibitem[{{Chambers} {et~al.}(2016){Chambers}, {Magnier}, {Metcalfe},
  {Flewelling}, {Huber}, {Waters}, {Denneau}, {Draper}, {Farrow}, {Finkbeiner},
  {Holmberg}, {Koppenhoefer}, {Price}, {Saglia}, {Schlafly}, {Smartt},
  {Sweeney}, {Wainscoat}, {Burgett}, {Grav}, {Heasley}, {Hodapp}, {Jedicke},
  {Kaiser}, {Kudritzki}, {Luppino}, {Lupton}, {Monet}, {Morgan}, {Onaka},
  {Stubbs}, {Tonry}, {Banados}, {Bell}, {Bender}, {Bernard}, {Botticella},
  {Casertano}, {Chastel}, {Chen}, {Chen}, {Cole}, {Deacon}, {Frenk},
  {Fitzsimmons}, {Gezari}, {Goessl}, {Goggia}, {Goldman}, {Grebel}, {Hambly},
  {Hasinger}, {Heavens}, {Heckman}, {Henderson}, {Henning}, {Holman}, {Hopp},
  {Ip}, {Isani}, {Keyes}, {Koekemoer}, {Kotak}, {Long}, {Lucey}, {Liu},
  {Martin}, {McLean}, {Morganson}, {Murphy}, {Nieto-Santisteban}, {Norberg},
  {Peacock}, {Pier}, {Postman}, {Primak}, {Rae}, {Rest}, {Riess}, {Riffeser},
  {Rix}, {Roser}, {Schilbach}, {Schultz}, {Scolnic}, {Szalay}, {Seitz},
  {Shiao}, {Small}, {Smith}, {Soderblom}, {Taylor}, {Thakar}, {Thiel},
  {Thilker}, {Urata}, {Valenti}, {Walter}, {Watters}, {Werner}, {White},
  {Wood-Vasey}, \& {Wyse}}]{Chambers}
{Chambers}, K.~C., {Magnier}, E.~A., {Metcalfe}, N., {et~al.} 2016, ArXiv
  e-prints, arXiv:1612.05560

\bibitem[{{{\'C}uk}(2018)}]{Cuk}
{{\'C}uk}, M. 2018, \apjl, 852, L15

\bibitem[{{Davidsson}(1999)}]{Davidsson}
{Davidsson}, B.~J.~R. 1999, \icarus, 142, 525

\bibitem[{{Debes} \& {Sigurdsson}(2002)}]{Debes}
{Debes}, J.~H., \& {Sigurdsson}, S. 2002, \apj, 572, 556

\bibitem[{{Debes} {et~al.}(2012){Debes}, {Walsh}, \& {Stark}}]{Debes2012}
{Debes}, J.~H., {Walsh}, K.~J., \& {Stark}, C. 2012, \apj, 747, 148

\bibitem[{{Do} {et~al.}(2018){Do}, {Tucker}, \& {Tonry}}]{Do}
{Do}, A., {Tucker}, M.~A., \& {Tonry}, J. 2018, \apjl, 855, L10

\bibitem[{{Dobrovolskis}(1990)}]{Dobrovolskis}
{Dobrovolskis}, A.~R. 1990, \icarus, 88, 24

\bibitem[{{Dohnanyi}(1969)}]{Doh}
{Dohnanyi}, J.~S. 1969, \jgr, 74, 2531

\bibitem[{{Drahus} {et~al.}(2017){Drahus}, {Guzik}, {Waniak}, {Handzlik},
  {Kurowski}, \& {Xu}}]{Drahus}
{Drahus}, M., {Guzik}, P., {Waniak}, W., {et~al.} 2017, ArXiv e-prints,
  arXiv:1712.00437

\bibitem[{{Elsanhoury} {et~al.}(2015){Elsanhoury}, {Nouh}, \&
  {Abdel-Rahman}}]{Elsan}
{Elsanhoury}, W.~H., {Nouh}, M.~I., \& {Abdel-Rahman}, H.~I. 2015, Revista
  Mexicana de Astronomia y Astrofisica, 51, 199

\bibitem[{{Engelhardt} {et~al.}(2017){Engelhardt}, {Jedicke}, {Vere{\v s}},
  {Fitzsimmons}, {Denneau}, {Beshore}, \& {Meinke}}]{Engel}
{Engelhardt}, T., {Jedicke}, R., {Vere{\v s}}, P., {et~al.} 2017, \aj, 153, 133

\bibitem[{{Farihi}(2016)}]{Farihi2016}
{Farihi}, J. 2016, New Astronomy Reviews, 71, 9

\bibitem[{{Farihi} {et~al.}(2010){Farihi}, {Barstow}, {Redfield}, {Dufour}, \&
  {Hambly}}]{Farihi2010}
{Farihi}, J., {Barstow}, M.~A., {Redfield}, S., {Dufour}, P., \& {Hambly},
  N.~C. 2010, \mnras, 404, 2123

\bibitem[{{Fitzsimmons} {et~al.}(2018){Fitzsimmons}, {Snodgrass}, {Rozitis},
  {Yang}, {Hyland}, {Seccull}, {Bannister}, {Fraser}, {Jedicke}, \&
  {Lacerda}}]{Fitz}
{Fitzsimmons}, A., {Snodgrass}, C., {Rozitis}, B., {et~al.} 2018, Nature
  Astronomy, 2, 133

\bibitem[{{Frewen} \& {Hansen}(2014)}]{Frewen}
{Frewen}, S.~F.~N., \& {Hansen}, B.~M.~S. 2014, \mnras, 439, 2442

\bibitem[{{Fukugita} \& {Peebles}(2004)}]{Fukugita}
{Fukugita}, M., \& {Peebles}, P.~J.~E. 2004, \apj, 616, 643

\bibitem[{{Gaidos}(2017)}]{Gaidos}
{Gaidos}, E. 2017, ArXiv e-prints, arXiv:1712.06721

\bibitem[{{Hansen} \& {Zuckerman}(2017)}]{Hansen}
{Hansen}, B., \& {Zuckerman}, B. 2017, Research Notes of the American
  Astronomical Society, 1, 55

\bibitem[{{Hollands} {et~al.}(2018){Hollands}, {G{\"a}nsicke}, \&
  {Koester}}]{Hollands2018}
{Hollands}, M.~A., {G{\"a}nsicke}, B.~T., \& {Koester}, D. 2018, \mnras, 477,
  93

\bibitem[{{Housen} \& {Holsapple}(1990)}]{Housen}
{Housen}, K.~R., \& {Holsapple}, K.~A. 1990, \icarus, 84, 226

\bibitem[{{Jackson} {et~al.}(2018){Jackson}, {Tamayo}, {Hammond}, {Ali-Dib}, \&
  {Rein}}]{Jackson}
{Jackson}, A.~P., {Tamayo}, D., {Hammond}, N., {Ali-Dib}, M., \& {Rein}, H.
  2018, \mnras, arXiv:1712.04435

\bibitem[{{Jewitt} {et~al.}(2017){Jewitt}, {Luu}, {Rajagopal}, {Kotulla},
  {Ridgway}, {Liu}, \& {Augusteijn}}]{Jewitt}
{Jewitt}, D., {Luu}, J., {Rajagopal}, J., {et~al.} 2017, \apjl, 850, L36

\bibitem[{{Johansen} {et~al.}(2015){Johansen}, {Mac Low}, {Lacerda}, \&
  {Bizzarro}}]{Johansen}
{Johansen}, A., {Mac Low}, M.-M., {Lacerda}, P., \& {Bizzarro}, M. 2015,
  Science Advances, 1, 1500109

\bibitem[{{Johnson} {et~al.}(2010){Johnson}, {Aller}, {Howard}, \&
  {Crepp}}]{Johnson2010}
{Johnson}, J.~A., {Aller}, K.~M., {Howard}, A.~W., \& {Crepp}, J.~R. 2010,
  \pasp, 122, 905

\bibitem[{{Johnson} {et~al.}(2007){Johnson}, {Butler}, {Marcy}, {Fischer},
  {Vogt}, {Wright}, \& {Peek}}]{Johnson2007}
{Johnson}, J.~A., {Butler}, R.~P., {Marcy}, G.~W., {et~al.} 2007, \apj, 670,
  833

\bibitem[{{Jura}(2003)}]{Jura2003}
{Jura}, M. 2003, \apjl, 584, L91

\bibitem[{{Jura}(2011)}]{Jura2011}
---. 2011, \aj, 141, 155

\bibitem[{{Jura} \& {Xu}(2010)}]{JuraXu2010}
{Jura}, M., \& {Xu}, S. 2010, \aj, 140, 1129

\bibitem[{{Jura} \& {Xu}(2012)}]{JuraXu}
---. 2012, \aj, 143, 6

\bibitem[{{Kalirai} {et~al.}(2008){Kalirai}, {Hansen}, {Kelson}, {Reitzel},
  {Rich}, \& {Richer}}]{Kalirai}
{Kalirai}, J.~S., {Hansen}, B.~M.~S., {Kelson}, D.~D., {et~al.} 2008, \apj,
  676, 594

\bibitem[{{Katz} \& {Dong}(2012)}]{Katz}
{Katz}, B., \& {Dong}, S. 2012, ArXiv e-prints, arXiv:1211.4584

\bibitem[{{Klein} {et~al.}(2011){Klein}, {Jura}, {Koester}, \&
  {Zuckerman}}]{Klein}
{Klein}, B., {Jura}, M., {Koester}, D., \& {Zuckerman}, B. 2011, \apj, 741, 64

\bibitem[{{Komossa}(2015)}]{Komossa}
{Komossa}, S. 2015, Journal of High Energy Astrophysics, 7, 148

\bibitem[{{Kozai}(1962)}]{Kozai}
{Kozai}, Y. 1962, \aj, 67, 591

\bibitem[{{Lacy} {et~al.}(1982){Lacy}, {Townes}, \& {Hollenbach}}]{Lacy}
{Lacy}, J.~H., {Townes}, C.~H., \& {Hollenbach}, D.~J. 1982, \apj, 262, 120

\bibitem[{{Laughlin} \& {Batygin}(2017)}]{Batygin}
{Laughlin}, G., \& {Batygin}, K. 2017, Research Notes of the American
  Astronomical Society, 1, 43

\bibitem[{{Lidov}(1962)}]{Lidov}
{Lidov}, M.~L. 1962, \planss, 9, 719

\bibitem[{{Lodato} {et~al.}(2009){Lodato}, {King}, \& {Pringle}}]{Lodato}
{Lodato}, G., {King}, A.~R., \& {Pringle}, J.~E. 2009, \mnras, 392, 332

\bibitem[{{Malamud} \& {Perets}(2017{\natexlab{a}})}]{Malamud1}
{Malamud}, U., \& {Perets}, H.~B. 2017{\natexlab{a}}, \apj, 842, 67

\bibitem[{{Malamud} \& {Perets}(2017{\natexlab{b}})}]{Malamud2}
---. 2017{\natexlab{b}}, \apj, 849, 8

\bibitem[{{Mamajek}(2017)}]{Mamajek}
{Mamajek}, E. 2017, Research Notes of the American Astronomical Society, 1, 21

\bibitem[{{McGlynn} \& {Chapman}(1989)}]{McGlynn}
{McGlynn}, T.~A., \& {Chapman}, R.~D. 1989, \apjl, 346, L105

\bibitem[{{Meech} {et~al.}(2017){Meech}, {Weryk}, {Micheli}, {Kleyna},
  {Hainaut}, {Jedicke}, {Wainscoat}, {Chambers}, {Keane}, {Petric}, {Denneau},
  {Magnier}, {Berger}, {Huber}, {Flewelling}, {Waters}, {Schunova-Lilly}, \&
  {Chastel}}]{Meech}
{Meech}, K.~J., {Weryk}, R., {Micheli}, M., {et~al.} 2017, \nat, 552, 378

\bibitem[{{Metzger} {et~al.}(2012){Metzger}, {Rafikov}, \&
  {Bochkarev}}]{Metzger}
{Metzger}, B.~D., {Rafikov}, R.~R., \& {Bochkarev}, K.~V. 2012, \mnras, 423,
  505

\bibitem[{{Morrison} \& {Malhotra}(2015)}]{Morrison}
{Morrison}, S., \& {Malhotra}, R. 2015, \apj, 799, 41

\bibitem[{{Movshovitz} {et~al.}(2012){Movshovitz}, {Asphaug}, \&
  {Korycansky}}]{Movsho}
{Movshovitz}, N., {Asphaug}, E., \& {Korycansky}, D. 2012, \apj, 759, 93

\bibitem[{{Mustill} \& {Villaver}(2012)}]{Mustill}
{Mustill}, A.~J., \& {Villaver}, E. 2012, \apj, 761, 121

\bibitem[{{Mustill} {et~al.}(2018){Mustill}, {Villaver}, {Veras},
  {G{\"a}nsicke}, \& {Bonsor}}]{Mustill2017}
{Mustill}, A.~J., {Villaver}, E., {Veras}, D., {G{\"a}nsicke}, B.~T., \&
  {Bonsor}, A. 2018, \mnras, 476, 3939

\bibitem[{{Petrovich} \& {Mu{\~n}oz}(2017)}]{PetMun}
{Petrovich}, C., \& {Mu{\~n}oz}, D.~J. 2017, \apj, 834, 116

\bibitem[{{Petrovich} {et~al.}(2014){Petrovich}, {Tremaine}, \&
  {Rafikov}}]{Petrovich}
{Petrovich}, C., {Tremaine}, S., \& {Rafikov}, R. 2014, \apj, 786, 101

\bibitem[{{Phinney}(1989)}]{Phinney}
{Phinney}, E.~S. 1989, in IAU Symposium, Vol. 136, The Center of the Galaxy,
  ed. M.~{Morris}, 543

\bibitem[{{Portegies Zwart} {et~al.}(2017){Portegies Zwart}, {Pelupessy},
  {Bedorf}, {Cai}, \& {Torres}}]{Portegies}
{Portegies Zwart}, S., {Pelupessy}, I., {Bedorf}, J., {Cai}, M., \& {Torres},
  S. 2017, ArXiv e-prints, arXiv:1711.03558

\bibitem[{{Rafikov}(2011{\natexlab{a}})}]{Rafikov1}
{Rafikov}, R.~R. 2011{\natexlab{a}}, \apjl, 732, L3

\bibitem[{{Rafikov}(2011{\natexlab{b}})}]{Rafikov2}
---. 2011{\natexlab{b}}, \mnras, 416, L55

\bibitem[{{Rafikov} \& {Silsbee}(2015)}]{RS15}
{Rafikov}, R.~R., \& {Silsbee}, K. 2015, \apj, 798, 70

\bibitem[{{Raymond} {et~al.}(2018){Raymond}, {Armitage}, {Veras}, {Quintana},
  \& {Barclay}}]{Raymond}
{Raymond}, S.~N., {Armitage}, P.~J., {Veras}, D., {Quintana}, E.~V., \&
  {Barclay}, T. 2018, \mnras, 476, 3031

\bibitem[{{Raymond} {et~al.}(2012){Raymond}, {Armitage}, {Moro-Mart{\'{\i}}n},
  {Booth}, {Wyatt}, {Armstrong}, {Mandell}, {Selsis}, \& {West}}]{Raymond2012}
{Raymond}, S.~N., {Armitage}, P.~J., {Moro-Mart{\'{\i}}n}, A., {et~al.} 2012,
  \aap, 541, A11

\bibitem[{{Rees}(1988)}]{Rees}
{Rees}, M.~J. 1988, \nat, 333, 523

\bibitem[{{Reffert} {et~al.}(2015){Reffert}, {Bergmann}, {Quirrenbach},
  {Trifonov}, \& {K{\"u}nstler}}]{Reffert}
{Reffert}, S., {Bergmann}, C., {Quirrenbach}, A., {Trifonov}, T., \&
  {K{\"u}nstler}, A. 2015, \aap, 574, A116

\bibitem[{{Sch{\"a}fer} {et~al.}(2017){Sch{\"a}fer}, {Yang}, \&
  {Johansen}}]{Schafer}
{Sch{\"a}fer}, U., {Yang}, C.-C., \& {Johansen}, A. 2017, \aap, 597, A69

\bibitem[{{Silsbee} \& {Rafikov}(2015)}]{Silsbee}
{Silsbee}, K., \& {Rafikov}, R.~R. 2015, \apj, 808, 58

\bibitem[{{Simon} {et~al.}(2016){Simon}, {Armitage}, {Li}, \&
  {Youdin}}]{Simon2016}
{Simon}, J.~B., {Armitage}, P.~J., {Li}, R., \& {Youdin}, A.~N. 2016, \apj,
  822, 55

\bibitem[{{Simon} {et~al.}(2017){Simon}, {Armitage}, {Youdin}, \&
  {Li}}]{Simon2017}
{Simon}, J.~B., {Armitage}, P.~J., {Youdin}, A.~N., \& {Li}, R. 2017, \apjl,
  847, L12

\bibitem[{{Sridhar} \& {Tremaine}(1992)}]{Sridhar}
{Sridhar}, S., \& {Tremaine}, S. 1992, \icarus, 95, 86

\bibitem[{{Stephan} {et~al.}(2017){Stephan}, {Naoz}, \& {Zuckerman}}]{Stephan}
{Stephan}, A.~P., {Naoz}, S., \& {Zuckerman}, B. 2017, \apjl, 844, L16

\bibitem[{{Stewart} \& {Leinhardt}(2009)}]{Stewart}
{Stewart}, S.~T., \& {Leinhardt}, Z.~M. 2009, \apjl, 691, L133

\bibitem[{{Stone} {et~al.}(2013){Stone}, {Sari}, \& {Loeb}}]{Stone}
{Stone}, N., {Sari}, R., \& {Loeb}, A. 2013, \mnras, 435, 1809

\bibitem[{{Trilling} {et~al.}(2017){Trilling}, {Robinson}, {Roegge},
  {Chandler}, {Smith}, {Loeffler}, {Trujillo}, {Navarro-Meza}, \&
  {Glaspie}}]{Trilling}
{Trilling}, D.~E., {Robinson}, T., {Roegge}, A., {et~al.} 2017, \apjl, 850, L38

\bibitem[{{Vanderburg} {et~al.}(2015){Vanderburg}, {Johnson}, {Rappaport},
  {Bieryla}, {Irwin}, {Lewis}, {Kipping}, {Brown}, {Dufour}, {Ciardi}, {Angus},
  {Schaefer}, {Latham}, {Charbonneau}, {Beichman}, {Eastman}, {McCrady},
  {Wittenmyer}, \& {Wright}}]{Vander}
{Vanderburg}, A., {Johnson}, J.~A., {Rappaport}, S., {et~al.} 2015, \nat, 526,
  546

\bibitem[{{Veras} {et~al.}(2014{\natexlab{a}}){Veras}, {Leinhardt}, {Bonsor},
  \& {G{\"a}nsicke}}]{Veras_TDE}
{Veras}, D., {Leinhardt}, Z.~M., {Bonsor}, A., \& {G{\"a}nsicke}, B.~T.
  2014{\natexlab{a}}, \mnras, 445, 2244

\bibitem[{{Veras} {et~al.}(2015){Veras}, {Leinhardt}, {Eggl}, \&
  {G{\"a}nsicke}}]{Veras_shrink}
{Veras}, D., {Leinhardt}, Z.~M., {Eggl}, S., \& {G{\"a}nsicke}, B.~T. 2015,
  \mnras, 451, 3453

\bibitem[{{Veras} {et~al.}(2014{\natexlab{b}}){Veras}, {Shannon}, \&
  {G{\"a}nsicke}}]{Veras2014}
{Veras}, D., {Shannon}, A., \& {G{\"a}nsicke}, B.~T. 2014{\natexlab{b}},
  \mnras, 445, 4175

\bibitem[{{Villaver} {et~al.}(2014){Villaver}, {Livio}, {Mustill}, \&
  {Siess}}]{Villaver}
{Villaver}, E., {Livio}, M., {Mustill}, A.~J., \& {Siess}, L. 2014, \apj, 794,
  3

\bibitem[{{Wyatt} {et~al.}(2014){Wyatt}, {Farihi}, {Pringle}, \&
  {Bonsor}}]{Wyatt2014}
{Wyatt}, M.~C., {Farihi}, J., {Pringle}, J.~E., \& {Bonsor}, A. 2014, \mnras,
  439, 3371

\bibitem[{{Wyatt} {et~al.}(2007){Wyatt}, {Smith}, {Su}, {Rieke}, {Greaves},
  {Beichman}, \& {Bryden}}]{Wyatt2007}
{Wyatt}, M.~C., {Smith}, R., {Su}, K.~Y.~L., {et~al.} 2007, \apj, 663, 365

\bibitem[{{Xu} {et~al.}(2017){Xu}, {Zuckerman}, {Dufour}, {Young}, {Klein}, \&
  {Jura}}]{Xu2017}
{Xu}, S., {Zuckerman}, B., {Dufour}, P., {et~al.} 2017, \apjl, 836, L7

\bibitem[{{Ye} {et~al.}(2017){Ye}, {Zhang}, {Kelley}, \& {Brown}}]{Ye}
{Ye}, Q.-Z., {Zhang}, Q., {Kelley}, M.~S.~P., \& {Brown}, P.~G. 2017, \apjl,
  851, L5

\bibitem[{{Zuckerman} \& {Young}(2017)}]{Zuckerman}
{Zuckerman}, B., \& {Young}, E.~D. 2017, ArXiv e-prints, arXiv:1707.03064

\end{thebibliography}

\appendix


\section{Useful formulae}  
\label{sect:formulae}


Here we collect explicit expressions for various functions mentioned throughout the text:
\ba  
\tilde f_2(z) &=& (2-\alpha)z^{\alpha-2}\left[
\frac{z^{2-\alpha}-1}{2(2-\alpha)}
-\frac{3\left(z^{1-\alpha}-1\right)}{4(1-\alpha)}
-\frac{z^{-(1+\alpha)}-1}{4(1+\alpha)}
\right],~~~~z>1,
\label{eq:f2}\\
\tilde f_3(z) &=& (4-\beta)z^{\beta-4}\int_1^z x^{3-\beta}\tilde f_2(x)dx,~~~~z>1.
\label{eq:f3}
\ea  

\end{document}